%% file: full_version.tex
\documentclass[letterpaper, 10pt, conference]{ieeeconf}      

\IEEEoverridecommandlockouts                
\usepackage{graphics} 
\usepackage{epsfig} 
\usepackage{times} 
\usepackage{amsmath} 
\usepackage{amssymb}  
\usepackage{psfrag}
\usepackage{subfigure}
\usepackage{mathptmx}
\usepackage{xcolor}
\usepackage{enumerate}
\usepackage{tikz}
\usepackage{pgfplots}
\usepackage{paralist}
\usepackage[]{algorithm2e}
\usepackage{arydshln}

\usetikzlibrary{arrows,shapes,trees}

\usepackage{nicefrac} 
\newtheorem{Theorem}{Theorem} 
\newtheorem{Lemma}{Lemma} 
\newtheorem{Remark}{Remark} 

\newcommand{\mc}[1]{\mathcal{#1}}

\newcommand{\ik}{^{(k)}}
\newcommand{\ikt}{^{(k)\top}}

\title{\LARGE \bf
Cooperative $\mathcal{H}_\infty-$Estimation \\ for Large-Scale Interconnected
Linear Systems%
\thanks{This work was supported by the German
Research Foundation (DFG) through
the Cluster of Excellence in Simulation Technology (EXC 310/1) at the
University of Stuttgart and the Australian Research Council under Discovery Projects funding scheme (Project DP120102152).}}

\author{
Jingbo Wu$^*$ \and Valery Ugrinovskii \and Frank Allg\"ower
\thanks{$^*$Corresponding author. 
J.~Wu and F.~Allg\"ower are with the Institute for Systems Theory and
Automatic Control, University of Stuttgart, Stuttgart, Germany. Email: {\tt\footnotesize \{jingbo.wu, allgower\}@ist.uni-stuttgart.de}. 
V.~Ugrinovskii is with the School of Information Technology and Electrical
Engineering, University of New South Wales at the 
Australian Defence Force Academy,
Canberra, Australia, Email: {\tt\footnotesize v.ougrinovski@adfa.edu.au}.}}

\begin{document}

\maketitle

\begin{abstract}
In this paper, a synthesis method for distributed estimation is presented,
which is suitable for dealing with large-scale interconnected linear systems with disturbance. The main feature of the proposed method is that local
estimators only estimate a reduced set of state variables and their 
complexity does not increase with the size of the system. Nevertheless, the local estimators are able to deal with lack of local detectability. Moreover, the
estimators guarantee $\mathcal{H}_\infty$-performance of the estimates with respect
to model and measurement disturbances. 
\medskip
\end{abstract}

\vspace{-0.5cm}
\section{Introduction}

Estimator design has been an essential part of controller design ever since
the development of state-space based controllers. A milestone was laid by
the Kalman Filter in 1960 \cite{Kalman1960}. 
While in the classical estimator design one estimator is used for the entire
system, decentralized and distributed estimators have gained attention
since decentralized control became one of the mainstream topics of interest
in control theory \cite{Corfmat1976,Speyer1979}. More results on
decentralized estimation were presented in
\cite{Siljak,Jang1994,Chung1995,Stankovic2009}. In decentralized estimation, typically,
a set of estimators are employed to create estimates of local subsystem
states with only limited assistance from each other. Couplings between
those subsystems are treated as undesirable disturbances. An
important requirement of this approach therefore is that the local subsystems
are detectable from local measurements. 


On the other hand, distributed filtering techniques like Distributed Kalman
Filtering were presented in
\cite{OlfatiSaber:2005vm,Olfati-saber2006,OlfatiSaber2007,Carli2008}. In a
distributed estimation setup, multiple estimators create an estimate of the
system's state, while cooperating with each other. 
Situations are not uncommon where individual estimators are unable
to obtain an estimate of the state on its own and cooperation becomes an
essential prerequisite \cite{Ugrinovskii2011,Ugrinovskii2011a}. At the same
time, the progress in the area of distributed estimation put forward issues
of scalability of estimator networks, i.e., there is an interest in
distributed estimation methods where the 
dimension of the local estimators does not increase with the total size of the
system. For instance, this is relevant for multi-agent systems, where the
agents are not able to perform a self-measurement, but only receive
relative information  \cite{Listmann2011,Wu2012}. 
Direct applications of the existing distributed estimation algorithms such
as those reported in
\cite{OlfatiSaber:2005vm,Olfati-saber2006,Ugrinovskii2011,Ugrinovskii2011a}
result in the estimators reproducing the entire state of the complete
network, and therefore, the order of the estimators grows with the size of the
network. However, for cooperative control of multi-agent systems, only
local information is required for each agent, making the estimate of the
whole network superfluous. This makes the direct application of the
mentioned distributed estimation algorithms ineffective with respect to the
necessary computation power. 

In this paper, our main contribution is the development of a new framework
that combines the benefits of both decentralized and distributed
estimation. 
We develop an estimation setup, where local estimators only
reproduce a desired subset of state variables and their complexity does not
grow with the total size of the system, in contrast to the existing methods
for distributed estimation mentioned above. Moreover, cooperation between
the local estimators will be used to deal with possible lack of local
detectability, which is a major difference to existing results
\cite{Siljak,Stankovic2009}. Therefore, the proposed setup is referred to as  
\emph{Cooperative Estimation}. In particular, we present an
$\mathcal{H}_\infty$-based design, which in addition provides guaranteed
performance with respect to model and measurement disturbances. 

This paper significantly extends our results in \cite{Wu2014} by
allowing a more general class of systems. In \cite{Wu2014}, we consider
dynamically decoupled agents and take
relative output measurements. Here, we expand the methodology to general large-scale linear systems, where
subsystems may be physically interconnected. This class of
systems is of high relevance due to its wide
range of applications, such as flexible structures \cite{Lynch2002} and electrical power
grids \cite{Schuler2014}. 

The paper is organized as follows: In Section 2, we present
some mathematical preliminaries and
the system class under consideration. Section 3 presents the
methodology of cooperative estimation design with 
guaranteed $\mathcal{H}_\infty$-performance, where the first step is an algorithm for
a judicious partition of the state space. 
In order to keep the presentation
simple and focused, we make the simplifying assumption that communications 
between the estimators are perfect. While perturbed communication channels 
may better reflect real-life applications,
this case can indeed also be considered within this estimation
setup although the resulting LMI conditions become more cumbersome;
cf.~\cite{Ugrinovskii2013}. 
Section 4 illustrates our results with a simulation
example
and Section 5 concludes the paper, and gives an overview on future
work. 

\section{Preliminaries}

Throughout the paper the following notation is used: 
Let $A$ be a quadratic matrix. If $A$ is positive definite, it is denoted
$A > 0$, and we write $A < 0$, if $A$ is negative definite. $0$ denotes a
matrix of suitable dimension, with all entries equal $0$. 

\subsection{Communication graphs}

In this section we summarize some notation from the graph theory.
We use directed, unweighted graphs $\mc{G} = (\mc{V} , \mc{E})$ to describe the communication topology between the individual agents. 
$\mc{V} = \{v_1,...,v_N\}$ is the set of vertices, where $v_k \in \mc{V}$ represents the $k$-th agent. 
$\mc{E} \subseteq \mc{V} \times \mc{V}$ is the set of edges, which model the information flow, i.e. the $k$-th agent receives information from agent $j$ if and only if $(v_j,v_k) \in \mc{E}$.
The set of vertices that agent $k$ receives information from is called the neighborhood of agent $k$, which is denoted by $\mathcal{N}\ik=\{j: (v_j,v_k) \in \mc{E}\}$. The outdegree of a vertex $k$ is defined as the number of edges in $\mc{E}$, which have $v_k$ as their tail. 



\subsection{System model}
We consider a large-scale linear time-invariant system, which consists of $N$ interconnected subsystems that are each described by the differential equation
\begin{align}
\dot{x}_k &= A_k x_k + \sum_{j =1}^N A_{kj} x_j + B_k v, \label{sys:LTI} \\
y_k  &= C_k x_k + \sum_{j=1}^N C_{kj} x_j + \eta_k, \label{sys:LTI_output}
\end{align}
for $k=1,...,N$, where $x_k \in \mathbb{R}^{n_k}$ is the state variable, $y_k \in \mathbb{R}^{r_k}$ is the output, and $v(t), \eta_k(t) \in
\mathcal{L}_2[0, \infty)$ are $\mathcal{L}_2$-integrable disturbance inputs of subsystem $k$. 
The scalar components of $x_k$ will be denoted $x_{k,i}$.
Note that in this system, all subsystems are
affected by the common disturbance $v$. This assumption does not lead to
loss of generality, since it also captures the case where the subsystems are
affected by different disturbances $v_k$, by simply stacking $v_k$ into one
vector.  

The global interconnected system can be written as
\begin{equation}\label{sys:lti_global}
\begin{aligned}
\dot{x} &= A x + B v, &
y  &= C x + \eta
\end{aligned}
\end{equation}
with
\begin{equation*}
\begin{aligned}
A&=\begin{bmatrix}
A_1 \!\!\!\! & A_{12} \!\!\!\! & \cdots \!\!\!\! & A_{1N} \\
A_{21} \!\!\!\! & A_2 \!\!\!\! &  \!\!\!\! & \vdots \\
\vdots \!\!\!\!& \!\!\!\!& \ddots \!\!\!\!& \vdots \\
A_{N1} \!\!\!\! & \hdots \!\!\!\!& \hdots \!\!\!\!& A_{N}
\end{bmatrix} &
C&= \begin{bmatrix} C_1 \!\!\!\! & C_{12} \!\!\!\! & \cdots \!\!\!\! & C_{1N} \\
C_{21} \!\!\!\! & C_2 \!\!\!\! &  \!\!\!\! & \vdots \\
\vdots \!\!\!\!& \!\!\!\!& \ddots \!\!\!\!& \vdots \\
C_{N1} \!\!\!\! & \hdots \!\!\!\!& \hdots \!\!\!\!& C_{N}
 \end{bmatrix} \\
 & B^\top=\begin{bmatrix}
 B_1^\top & \hdots & B_N^\top
 \end{bmatrix}
\end{aligned}
\end{equation*}
by using the stacked state and disturbance vectors $ x = [ x_1^\top , ...,
x_N^\top ]^\top \in \mathbb{R}^n $ and $\eta = [ \eta_1^\top ,...,
\eta_N^\top]^\top$. 

\medskip
\textbf{Assumption 1:} The global plant $(A,C)$ is observable.
\medskip

It is well
known that Assumption 1 is a sufficient condition in the centralized case. In this paper, this
assumption is to setup a basic framework under which the state estimation
problem under consideration is meaningful.   


%
%

\subsection{Re-partitioning of the system}

In the next section, we will re-partition the vector $x$ for designing
local estimators. Associated with the collection of outputs
\eqref{sys:LTI_output}, for every $k=1,...,N$, we choose a
$\sigma_k$-dimensional partial state vector  
\begin{equation}\label{sys:partition_xk}
x^{(k)} = \begin{bmatrix}
x_{\xi_k(1)} \\ \vdots \\ x_{\xi_k(\sigma_k)}
\end{bmatrix},
\end{equation}
where $\xi_k(\cdot)$ is a selection function that determines, which scalar components $x_{j,i}$ are included in $x\ik$. This represents a degree of freedom in the design of the estimators and all elements of the global state vector $x$ may be chosen that are relevant to subsystem $k$. For instance, $x\ik$ may contain $x_k$, but it does not have
to include all of them, if for subsystem $k$, some parts of its own state
are not important. In particular, it is required that all $x_{j,i}$ which contribute towards $y_k$ are included in $x\ik$. As a result, every output $y_k$ can be equivalently expressed as
\begin{equation}\label{sys:sensing_agent}
y_k = C\ik x\ik + \eta_k.
\end{equation}

One possible choice of $x\ik$ is the stacked
vector including $x_k$ and all $x_j$ with $C_{kj} \neq 0$. In that case,  
\begin{equation*}
A\ik = \begin{bmatrix}
A_k & A_{kj_1} & \hdots \\
A_{j_1k} & A_{j_1} &  \\
\vdots & & \ddots
\end{bmatrix}
\end{equation*}
and the rest of the coefficients in (\ref{sys:global_k}) are defined in a
similar fashion.

The selection function $\xi_k$ is a discrete injective map
\begin{equation}
\xi_k: \{1,...,\sigma_k\} \rightarrow \mathcal{Y}, \quad \sigma_k \leq n,
\end{equation}
where the set 
$\mathcal{Y} \triangleq \{ (k,i) | k=1,...,N; i=1,...,n_k\} $
is defined as the combination of all appearing indexes of the subsystem
states and their scalar components $x_{k,i}$. For the ease of notation, we
refer to the elements of the set $\mathcal{Y}$ as $\lambda$, i.e., $\lambda =
(k,i) \in \mathcal{Y}$.   
 
The image of $\xi_k$ is denoted as $I\ik$, $I\ik\subset \mathcal{Y}$, and
the inverse map $\xi_k^{-1}$ is an enumeration of the elements of $I\ik$, 
\begin{equation}
\xi_k^{-1} : I\ik \rightarrow \{1,...,\sigma_k\},
\end{equation}
which assigns a position in $x\ik$ to selected components $x_{\lambda}$ of
the global state vector $x$.

In general, partial state vectors $x^{(k)}$ may overlap, e.g. $x^{(1)}$ and $x^{(2)}$ may contain a common component $x_\lambda$.

For all $k=1,..,N$ the global interconnected system \eqref{sys:lti_global}
can now be written as
\begin{equation}\label{sys:global_k}
\begin{bmatrix}
\dot{x}\ik \\ \dot{x}\ik_c
\end{bmatrix} = \begin{bmatrix}
A\ik & \widetilde{A}\ik \\
\widetilde{A}\ik_c & A\ik_c
\end{bmatrix}
\begin{bmatrix}
x\ik \\ x\ik_c
\end{bmatrix} + \begin{bmatrix}
B\ik \\ B\ik_c
\end{bmatrix} v
\end{equation}
by permutation of the states.

For every $k$, the composition of the matrices $A\ik$, $B\ik$, etc., is
determined by the composition of the partial state variable $x\ik$; in
turn, the latter is determined by the components of the global state $x$
which estimator $k$ seeks to obtain (see Section~\ref{Section.Design}). 

\section{Cooperative Estimation Problem and the Estimator Design}\label{Section.Design}

The problem considered in this paper is to design a local estimator for
every subsystem $k$ that creates an estimate for the local partial state
variable $x\ik$ using the local measurements $y_k$ described in
(\ref{sys:sensing_agent}). The vector of local estimates will be denoted 
\begin{equation*}
\hat{x}^{(k)}= \begin{bmatrix}
\hat{x}\ik_{\xi_k(1)} \\
\vdots \\
\hat{x}\ik_{\xi_k(\sigma_k)}
\end{bmatrix} \in \mathbb{R}^{\sigma_k},
\end{equation*}
where $\hat{x}\ik_\lambda$ is the estimate for $x_\lambda$ computed at
subsystem $k$. 

The local estimation error vector is defined as
\begin{equation*}
\epsilon^{(k)}= x^{(k)} -\hat{x}^{(k)} =\begin{bmatrix}
x_{\xi_k(1)} - \hat{x}_{\xi_k(1)} \\
\vdots \\
x_{\xi_k(\sigma_k)} - \hat{x}_{\xi_k(\sigma_k)}
\end{bmatrix}.
\end{equation*}
We now formally pose the estimator synthesis problem. \medskip

\textbf{Problem 1:}
Determine a collection of estimates $\hat{x}^{(k)}(t)$, $k=1, \ldots,N$,  such
that the following two properties are satisfied simultaneously. 
\begin{compactenum}[(i)]
\item In the absence of model and measurement disturbances (i.e., when
  $v=0$, $\eta=0$), the estimation errors decay so that $\epsilon\ik \to 0$
  exponentially for all $k=1,...,N$. 

\item The estimators \eqref{eq:estimator} provide guaranteed
  $\mathcal{H}_{\infty}$ performance in the sense that 
\begin{equation}\label{eq:hinf-performance}
\begin{aligned}
&   \sum_{k=1}^N \int_0^\infty \epsilon^{(k)\top}W\ik \epsilon\ik  dt\\
&\leq  \sum_{k=1}^N \int_0^\infty \left(\omega^2 \| v \|^2 + \gamma^2 \| \eta_k \|^2 \right) dt + I_0,
\end{aligned}
\end{equation}
for a positive semi-definite weighting matrix $W\ik$. In (\ref{eq:hinf-performance}),
$I_0=\sum_{k=1}^N
x^{(k)\top}_0 P\ik x\ik_0$ is the cost due to the observer's uncertainty about
the initial conditions of the subsystems.  
\end{compactenum}

%
This will be achieved by allowing certain agents to communicate with each other. 

\subsection{Communication Requirements}

There are two factors, which influence the required communication for the
cooperative estimation setup: The first one is detectability of $(A\ik,
C\ik)$. In the special case of $\widetilde{A}\ik=0$, and $(A\ik, C\ik)$
being detectable for all $k=1,...,N$, no communication is necessary at all,
as for every subsystem, an estimator can be designed separately. However,
these assumptions may not hold in a general case. In particular, in this
paper, we do not require that $(A\ik, C\ik)$ are detectable for all
$k=1,...,N$, which is a major difference compared to existing methods in
literature, for instance, \cite{Siljak} and \cite{Stankovic2009}. In fact, even all $(A\ik, C\ik)$ may be undetectable.

The second factor which influences the required communication is sparsity
of $\widetilde{A}\ik$. Ideally, when the partial state $x\ik$ is decoupled
from the rest of the system, i.e. $\widetilde{A}\ik=0$, a standard
$\mathcal{H}_\infty$  filter can be employed to carry out the estimation of
$x\ik$ from $y_k$. However, if $x\ik$ includes a state $x_{\lambda}$, which
is connected to a state $x_{\lambda^*}$ that is not a component of $x\ik$,
then the problem becomes more challenging. When the connections strength is
limited, this can be handled using methods like the Small Gain
Theorem. Otherwise, communication may be used to compensate for coupling
between $x\ik$ and $x\ik_c$. In this paper, we investigate the latter
method. 

In order to define the required communication channels, we use an assignment function
\begin{equation}
\zeta: \mathcal{Y} \to \{0, 1,...,N\},
\end{equation}
with the property that
\begin{equation*}
\lambda \in I^{(\zeta(\lambda))},
\end{equation*}
if $\zeta(\lambda) \neq 0$. Moreover, $\zeta(\lambda) =0$ only if $\lambda
\not\in I\ik$ for all $k=1,...,N$. The map $\zeta(\cdot)$ assigns
responsibilities in estimating the system's states to the subsystems and
their local estimators. In general, $\zeta(\lambda)$ is not unique and
there is a degree of freedom in selection of the assignment function.   
However in the case when $x\ik$'s do not overlap, the assignment function
$\zeta(\lambda)$ is unique. 

With the definition of the assignment function $\zeta$, we can introduce the assumption on the communication graph used in this paper:

\medskip

\noindent\textbf{Assumption 2:} If a component $x_{\lambda}$ of $x\ik$ is physically coupled to a state $x_{\lambda^*}$, where $\lambda^* \in \mathcal{Y} \backslash I\ik$,
then subsystem $j=\zeta(\lambda^*) \neq 0$ can communicate to subsystem $k$, i.e. $(j,k) \in \mathcal{E}$. 
\medskip

We denote with $I\ik_c$ the set of all indexes $\lambda^* \in \mathcal{Y}
\backslash I\ik$  with the property that for all $x_{\lambda^*} \in I\ik_c$, there exists a component $x_\lambda$ of $x\ik$, which is coupled to $x_{\lambda^*}$. Assumption 2 reflects the point made
above, as the more entries $\widetilde{A}\ik$ has, the more communication
between the subsystems is required. Some remarks on the realization of this
assumption are in order: 

\begin{Remark}
In the literature (see e.g. \cite{Siljak}, \cite{Lynch2002}), it is often assumed that the
communication topology simply mimics the interconnection topology. Assumption 2
is more general in this respect as the selection of the assignment function
$\zeta$ and the 
definition of the partition $x\ik$ imply a certain degree of
freedom. In particular, if the partial state $x\ik$ is coupled to some
state $x_{\lambda^*}$ which is not a component of $x\ik$, there are two
ways for subsystem $k$ to obtain the required information: 
\begin{enumerate}
\item One intuitive option for subsystem $k$ is to receive an
  estimate of $x_{\lambda^*}$ through communication with another
  subsystem. The transmitting subsystem  
may be any subsystem, which is assigned to estimate $x_{\lambda^*}$ by the
choice of the assignment function $\zeta$. The function $\zeta$ should
therefore be selected so that subsystem $j=\zeta(\lambda^*)$ can
communicate to subsystem $k$. 
\item An alternative option is to include $x_{\lambda^*}$ into the partial
  state $x\ik$, and therefore, assign the task of estimating
  $x_{\lambda^*}$ to subsystem $k$. This circumvents the communication
  requirement of Assumption 2. This method is meaningful, when the local
  estimator has enough computational power to handle additional coordinates.
\end{enumerate}
\end{Remark}

\begin{Remark}
In some applications, the communication topology is not a fixed system property, but is a design parameter. In that case, Assumption 2 can be easily realized.
\end{Remark}



\medskip
\begin{Lemma}
For all $\lambda \in \mathcal{Y}$ there exists a $k \in \{1,...,N\}$, such that $\lambda \in I\ik$.

\proof{Suppose there exists a $\lambda \in \mathcal{Y}$, such that $\lambda \not\in I\ik$ for all $k=1,...,N$. By the definition of the partial states $x\ik$ and the selection function $\xi_k$, the column of $C$ which corresponds to $x_\lambda$ is $0$. Moreover, by the definition of $\zeta$, we have $\zeta(\lambda)=0$ and thus, it follows from Assumption 2 that there is no partial state $x\ik$ that is coupled to $x_\lambda$. Therefore, $x_\lambda$ is not observable, which contradicts Assumption 1.
}
\end{Lemma}

\medskip

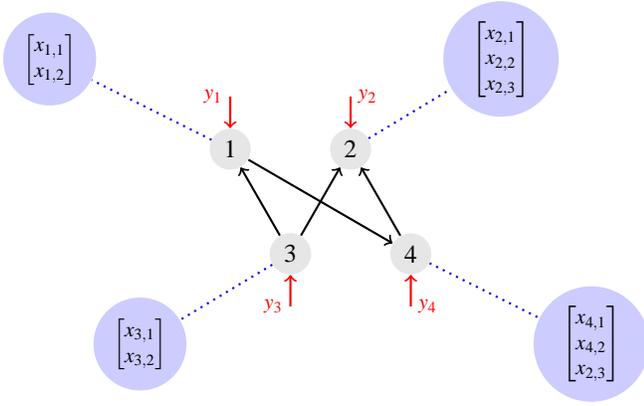
\begin{figure} 
\begin{center}
\setlength{\unitlength}{1mm}
\begin{tikzpicture}[scale=0.8, transform shape]
\node [circle, fill=gray!20] (a) at (0,2*0.87) {\large $1$};
\node (y_1) at (0,2*0.87+1) {};
\node [circle, fill=blue!20] (a_est) at (-3,2*0.87+1.5) {$\begin{bmatrix}
x_{1,1} \\
x_{1,2}
\end{bmatrix}$};
\node [circle, fill=gray!20] (b) at (2,2*0.87) {\large $2$};
\node (y_2) at (2,2*0.87+1) {};
\node [circle, fill=blue!20] (b_est) at (4.5,2*0.87+1.5) {$\begin{bmatrix}
x_{2,1} \\
x_{2,2} \\
x_{2,3}
\end{bmatrix}$};
\node [circle, fill=gray!20] (c) at (1,0) {\large $3$};
\node (y_3) at (1,-1) {};
\node [circle, fill=blue!20] (c_est) at (-1.5,-1.5) {$\begin{bmatrix}
x_{3,1} \\
x_{3,2}
\end{bmatrix}$};
\node [circle, fill=gray!20] (d) at (3,0) {\large $4$};
\node (y_4) at (3,-1) {};
\node [circle, fill=blue!20] (d_est) at (6,-1.5) {$\begin{bmatrix}
x_{4,1} \\
x_{4,2} \\
x_{2,3} 
\end{bmatrix}$};
\draw [thick, ->] (a) -- (d);
\draw [thick, ->] (c) -- (a);
\draw [thick, ->] (c) -- (b);
\draw [thick, ->] (d) -- (b);
\draw [thick, dotted, blue] (a) -- (a_est);
\draw [thick, dotted, blue] (b) -- (b_est);
\draw [thick, dotted, blue] (c) -- (c_est);
\draw [thick, dotted, blue] (d) -- (d_est);
\draw [thick, red, ->] (y_1) -- (a)node [pos=0, left] {$y_1$};;
\draw [thick, red, ->] (y_2) -- (b)node [pos=0, right] {$y_2$};;
\draw [thick, red, ->] (y_3) -- (c)node [pos=0, left] {$y_3$};;
\draw [thick, red, ->] (y_4) -- (d)node [pos=0, right] {$y_4$};;

\end{tikzpicture}
\caption{Example of an admissible estimator structure. The nodes in the
  center represent the subsystems, which receive measurement information
  and estimate the partial states $x\ik$ that are written in the outer circles. The
  black arrows represent the communication links between the subsystems. 
}
\label{fig:estimator_setup}
\end{center}
\end{figure}

An example for the interconnection structure  which satisfies Assumption 2
is shown in Figure \ref{fig:estimator_setup}. This communication structure
pertains to the numerical example presented later in Section IV. 
As noted for the partition \eqref{sys:partition_xk}, the vectors $x\ik$ may
overlap. Therefore, including a consensus term whenever overlapping estimators can communicate is able to enhance estimation performance of the
subsystems and in some cases even facilitates feasibility of the design
conditions which will be introduced later in the paper. 

\subsection{Formal definition of estimators}

The estimator dynamics are now proposed for each subsystem as
\begin{equation}\label{eq:estimator}
\begin{aligned}
\dot{\hat{x}}^{(k)} =& A\ik \hat{x}^{(k)} + L\ik(y_k-C\ik \hat{x}\ik) + \sum_{\lambda \in I\ik_c} \left[\widetilde{A}\ik \right]_{\lambda} \hat{x}^{(\zeta(\lambda))}_{\lambda} \\
& +K\ik \sum_{j \in \mathcal{N}_k} \left( \sum_{\lambda \in I\ik \cap I^{(j)}}
e_{\xi^{-1}(\lambda)}(\hat{x}^{(j)}_{\lambda}-  \hat{x}\ik_{\lambda}) \right)
\end{aligned}
\end{equation}
with initial condition $\hat{x}^{(k)}_0 = 0$, where $\left[\widetilde{A}\ik \right]_{\lambda}$ is the column of $\widetilde{A}\ik$ which corresponds to $x_{\lambda}$ and the unit vector $e_{\xi^{-1}(\lambda)}$ injects the difference $\hat{x}^{(l)}_{\lambda}-  \hat{x}\ik_{\lambda}$ to the $\sigma_k$-dimensional space.



Problem 1 can now be particularized using the
estimators~(\ref{eq:estimator}).  
\medskip

\textbf{Problem 1':}
Determine estimator gains $L\ik$, $K\ik$ in \eqref{eq:estimator} such that
properties (i) and (ii) of Problem 1 hold.

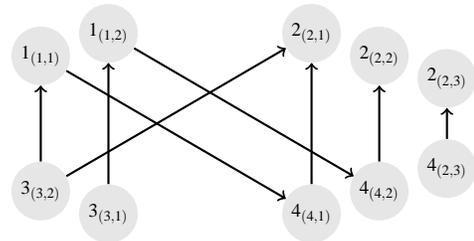
\begin{figure}[b] 
\begin{center}
\setlength{\unitlength}{1mm}
\begin{tikzpicture}[scale=0.6, transform shape]
\node [circle, fill=gray!20] (a_1) at (0,2) {\large $1_{(1,1)}$};
\node [circle, fill=gray!20] (a_2) at (1.5,2.5) {\large $1_{(1,2)}$};
\node [circle, fill=gray!20] (b_3) at (6,2.5) {\large $2_{(2,1)}$};
\node [circle, fill=gray!20] (b_4) at (7.5,2) {\large $2_{(2,2)}$};
\node [circle, fill=gray!20] (b_5) at (9,1.5) {\large $2_{(2,3)}$};
\node [circle, fill=gray!20] (c_6) at (0,-1) {\large $3_{(3,2)}$};
\node [circle, fill=gray!20] (c_7) at (1.5,-1.5) {\large $3_{(3,1)}$};
\node [circle, fill=gray!20] (d_5) at (9,-0.5) {\large $4_{(2,3)}$};
\node [circle, fill=gray!20] (d_8) at (7.5,-1) {\large $4_{(4,2)}$};
\node [circle, fill=gray!20] (d_9) at (6,-1.5) {\large $4_{(4,1)}$};
\draw [thick, ->] (a_1) -- (d_9);
\draw [thick, ->] (a_2) -- (d_8);
\draw [thick, ->] (c_6) -- (b_3);
\draw [thick, ->] (c_7) -- (a_2);
\draw [thick, ->] (c_6) -- (a_1);
\draw [thick, ->] (d_5) -- (b_5);
\draw [thick, ->] (d_8) -- (b_4);
\draw [thick, ->] (d_9) -- (b_3);
\end{tikzpicture}
\caption{Example of the extended graph $\widetilde{\mathcal{G}}$: The subsystems are
  now represented by clusters of $\sigma_k$ vertices, where every vertex
  represents a single estimator coordinate. 
}
\label{fig:extended_graph}
\end{center}
\end{figure}

In order to solve this problem, we define the extended graph
$\widetilde{\mathcal{G}}$, which will be used in the analysis of the
interconnection structure between the subsystems. Let every subsystem be
represented by a cluster of $\sigma_k$ nodes, where vertex $v_{k_\lambda}$
represents the estimator state $\hat{x}\ik_\lambda$. The edges of
$\widetilde{\mathcal{G}}$ are now determined by Algorithm
\ref{alg:extendedgraph}. An example for the extended graph is shown in
Figure \ref{fig:extended_graph}, which applies to the numerical
example presented in Section IV. 
\vspace{-0.3cm}
\begin{algorithm}\label{alg:extendedgraph}
\textbf{Algorithm 1} \\
\KwResult{Graph $\widetilde{\mathcal{G}}$}
Set the edge set $\widetilde{\mathcal{E}}=\{\}$.\\
\For{$k=1,...,N$}
{
\For{$\lambda^* \in I_c\ik$ (\textbf{\emph{interconnection part}})}{
Add the edge $(v_{\zeta(\lambda^*)_{\lambda^*}},v_{k_\lambda})$ to $\widetilde{\mathcal{E}}$, where $x_\lambda$ is the component of $x\ik$ that is coupled to $x_{\lambda^*}$. 
}
\For{$j \in \mc{N}_k$ and $\lambda \in I\ik \cap I^{(j)}$ (\textbf{\emph{fusion part}})}{
Add an edge $(v_{j_\lambda}, v_{k_\lambda})$ to $\widetilde{\mathcal{E}}$.
}
}
\end{algorithm}
\vspace{-0.3cm}

The graph generated by Algorithm 1 graphically displays the detailed connection structure of the estimation vectors $\hat{x}\ik$. The out-degree of vertex $v_{k_\lambda}$ in the extended graph is denoted
by $q(k,\lambda)$. This definition will be used to present our main
results on the design of the filter gains, which are given in the next
section. 


\subsection{Filter gains design}
We define the matrices
\begin{align*}
N\ik =& \sum_{j \in \mc{N}_k} \left( \sum_{\lambda \in I\ik \cap I^{(j)}} e_{\xi_k^{-1}(\lambda)} e^\top_{\xi_k^{-1}(\lambda)} \right) \\
Q\ik =& P\ik A\ik + A^{(k)\top} P\ik - G\ik C\ik - (G \ik C\ik )^\top \\ -& F\ik N\ik - (F\ik N\ik)^\top \\
+& \alpha P\ik + \underbrace{\pi_k \begin{bmatrix}
 q(k, \xi_k(1)) P\ik_{1} & 0 & 0 \\
 0 & \ddots & 0 \\
 0 & 0 & q(k, \xi_k(\sigma_k)) P\ik_{\sigma_k }
\end{bmatrix}}_{\Pi_k},
\end{align*}
where $G\ik \in \mathbb{R}^{\sigma_k \times r_k}$ and $F\ik \in \mathbb{R}^{\sigma_k \times \sigma_k}$ are unknown matrices, $P\ik \in \mathbb{R}^{\sigma_k \times \sigma_k}$ is a symmetric,
positive definite matrix and $P\ik_{i} \in \mathbb{R}$ is
the $i$-th diagonal element of $P\ik$. $\pi_k$ and $\alpha$ are positive 
constants which will later play the role of design parameters. Furthermore, we define $p(\lambda)$ as the diagonal element of $P^{(\zeta(\lambda))}$ which corresponds to $x_\lambda$, i.e. the $\xi^{-1}_{\zeta(\lambda)}(\lambda)$'th diagonal element.
%


Next, for all $k=1,...,N$, we define the matrices
\begin{align*}
S\ik &= \begin{bmatrix}
P\ik \left[\widetilde{A}\ik\right]_{\lambda\ik_1} & P\ik \left[\widetilde{A}\ik\right]_{\lambda\ik_2} & \hdots  
\end{bmatrix} \\
R\ik &= \begin{bmatrix} \pi_{\zeta(\lambda\ik_1)} p(\lambda\ik_1) & 0 & 0\\
0 &  \pi_{\zeta(\lambda\ik_2)} p(\lambda\ik_2) & 0\\
0 & 0 & \ddots  \\
\end{bmatrix}, 
\end{align*}
for $\{ \lambda\ik_1, \lambda\ik_2, ... \} = I\ik_c$ and
\begin{align*}
T_j\ik &= \begin{bmatrix} F\ik e_{\xi_k^{-1}(\lambda_{1,j})} & F\ik e_{\xi_k^{-1}(\lambda_{2,j})} & \hdots
\end{bmatrix}\\
U_j\ik &= \begin{bmatrix} \pi_{j} P^{(j)}_{\xi^{-1}_{j}(\lambda_{1}^{kj})} & 0 & 0 \\
0 & \pi_{j} P^{(j)}_{\xi^{-1}_{j}(\lambda_{2}^{kj})} & 0 \\
0 & 0 & \ddots
\end{bmatrix},
\end{align*}
where $\{\lambda_{1}^{kj}, \lambda_{2}^{kj}, ... \} = I\ik \cap I^{(j)}$.

With these definitions, we are ready to present following theorem. 

\begin{Theorem}
Consider a group of interconnected LTI systems \eqref{sys:LTI} with local outputs \eqref{sys:LTI_output}. 
Problem 1 admits a solution in the form of estimators (\ref{eq:estimator})
with 
\begin{equation}\label{eq:filtergains}
\begin{aligned}
L\ik &= (P^{(k)})^{-1} G\ik, \\
K\ik &= (P^{(k)})^{-1} F\ik,
\end{aligned}
\end{equation}
if the matrices $F\ik$, $G\ik$ and $P\ik$, $k=1,\ldots,N$ are a solution of the following LMIs:

\begin{equation}\label{LMI}
\begin{aligned}
\left[\begin{array}{ccc;{2pt/2pt}cccc}
\!\!\!\!\!\! Q\ik + W\ik & \!\!\! -G\ik & \!\!\! P\ik B\ik & S\ik & \!\!\! T_{j_{1,k}}\ik & \hdots & \!\!\! T_{j_{\tau_k,k}}\ik \\
- G^{(k)\top} & \!\!\! -\gamma^2 I & 0  & 0 & 0 & 0 & 0\\ 
\!\! (P\ik B\ik)^\top & 0 & \!\!\! -\omega^2 I  & 0 & 0 & 0 & 0\\
\hdashline[2pt/2pt]
S^{(k)\top} & 0 & 0 &  R\ik & 0 & 0 & 0\\
T_{j_{1,k}}^{(k)\top} & 0 & 0 &  0 & \!\!\! U_{j_{1,k}}\ik & 0 & 0 \\
\vdots & 0 & 0 &  0 & 0 & \ddots & 0 \\
T_{j_{\tau_k,k}}^{(k)\top} & 0 & 0 &  0 & 0 & 0 & \!\!\! U_{j_{\tau_k,k}}\ik
\end{array}\right] \!\!\!\!
< \!\! 0
\end{aligned}
\end{equation} 
with $\{j_{1,k}, j_{2,k}, ..., j_{\tau_k,k}\} = \mc{N}_k$.

\medskip

\proof{ The estimator error dynamics at node $k$ are
\begin{align*}
\dot{\epsilon}\ik =& (A\ik - L\ik C\ik )\epsilon\ik + \sum_{\lambda \in I\ik_c} \left[\widetilde{A}\ik \right]_\lambda \epsilon^{(\zeta(\lambda))}_\lambda \\
& +K\ik \sum_{j \in \mathcal{N}_k} \left(\sum_{\lambda \in I\ik \cap I^{(j)}}
e_{\xi^{-1}(\lambda)}(\epsilon^{(j)}_\lambda-  \epsilon\ik_\lambda) \right) \\
& - L\ik \eta\ik + B\ik v.
\end{align*}

We use a Lyapunov function
\begin{equation*}
V(\epsilon)= \sum_{k=1}^N \underbrace{\epsilon^{(k)\top}P^{(k)}\epsilon\ik}_{V\ik(\epsilon\ik)},
\end{equation*}
where $V\ik(\epsilon\ik)$ are the individual components of $V(\epsilon)$.

The Lie derivative of $V\ik(\epsilon\ik)$ is
\begin{align*}
\dot{V}\ik(\epsilon\ik) =& 2 \epsilon^{(k)\top} P\ik (A\ik - L\ik C\ik) \epsilon\ik \\
&+ 2 \epsilon^{(k)\top} P\ik(- L\ik \eta\ik + B\ik v) \\ 
&+ 2 \epsilon\ikt P\ik \sum_{\lambda \in I\ik_c} \left[\widetilde{A}\ik \right]_\lambda \epsilon^{(\zeta(\lambda))}_\lambda \\
&+ 2 \epsilon\ikt P\ik K\ik \sum_{j \in \mathcal{N}_k} \sum_{\lambda \in I\ik \cap I^{(j)}}
e_{\xi^{-1}(\lambda)}(\epsilon^{(j)}_\lambda-  \epsilon\ik_\lambda) \\
=& 2 \epsilon^{(k)\top} \! P\ik \!\!\! \left( \!\! A\ik \!\!-\!\! L\ik C\ik \!\!-\!\! K\ik \! N\ik \!\! \right) \!\! \epsilon\ik \\ 
&+ 2 \epsilon^{(k)\top} P\ik(- L\ik \eta\ik + B\ik v) \\ 
&+ 2 \epsilon\ikt P\ik \sum_{\lambda \in I\ik_c} \left[\widetilde{A}\ik \right]_\lambda \epsilon^{(\zeta(\lambda))}_\lambda \\
&+ 2 \epsilon\ikt P\ik K\ik \sum_{j \in \mathcal{N}_k} \sum_{\lambda \in I\ik \cap I^{(j)}}
e_{\xi^{-1}(\lambda)}\epsilon^{(j)}_\lambda
\end{align*}

With the filter gains \eqref{eq:filtergains}
and the LMIs \eqref{LMI} it can be obtained that
\begin{align*}
\dot{V}\ik(\epsilon) =& \epsilon^{(k)\top} \left( Q\ik - \alpha P\ik - \Pi_k \right)\epsilon\ik\\
&- 2 \epsilon^{(k)\top} G\ik \eta\ik + 2 e^{(k)\top} P\ik B\ik v \\ 
&+ 2 \epsilon\ikt P\ik \sum_{\lambda \in I\ik_c} \left[\widetilde{A}\ik \right]_\lambda \epsilon^{(\zeta(\lambda))}_\lambda \\
&+ 2 \epsilon\ikt F\ik \sum_{j \in \mathcal{N}_k} \sum_{\lambda \in I\ik \cap I^{(j)}}
e_{\xi^{-1}(\lambda)}\epsilon^{(j)}_\lambda \\
\leq & 
\sum_{\lambda \in I\ik_c} \epsilon^{(\zeta(\lambda))\top}_\lambda \pi_{\zeta(\lambda)} p(\lambda) \epsilon^{(\zeta(\lambda))}_\lambda \\
&+ \sum_{j \in \mathcal{N}_k} \sum_{\lambda \in I\ik \cap I^{(j)}} \epsilon^{(j)\top}_\lambda \pi_{j} P^{(j)}_{\xi^{-1}_{j}(\lambda)} \epsilon^{(j)}_\lambda \\
&- \epsilon^{(k)\top} W\ik \epsilon\ik +\gamma^2 \eta^{(k)\top}\eta\ik + \omega^2 v^\top v \\
&- \alpha \epsilon^{(k)\top} P\ik \epsilon\ik - \epsilon^{(k)\top} \Pi_k \epsilon^{(k)}.
\end{align*}
Summing up the $V\ik$s, it holds for $V$ that
\begin{equation*}
\begin{aligned}
\dot{V}(\epsilon) \leq & \sum_{k=1}^N \pi_{k} \sum_{\lambda \in I\ik} q(k,\lambda) \epsilon^{(k)T}_\lambda P^{(k)}_{\lambda} \epsilon^{(k)}_\lambda \\ 
& - \sum_{k=1}^N \epsilon^{(k)\top} W\ik \epsilon\ik + \sum_{k=1}^N\gamma^2 \eta^{(k)\top}\eta\ik + \sum_{k=1}^N\omega^2 v^\top v \\
&- \sum_{k=1}^N \alpha \epsilon^{(k)\top} P\ik \epsilon\ik -  \sum_{k=1}^N \epsilon^{(k)\top} \Pi_k \epsilon^{(k)}
\end{aligned}
\end{equation*}

\begin{equation}\label{eq:Vdot}
\begin{aligned}
\dot{V}(\epsilon) \leq& -\alpha \sum_{j=1}^N  \underbrace{\epsilon^{(k)\top} P\ik \epsilon\ik}_{V\ik} - \sum_{k=1}^N \epsilon^{(k)\top} W\ik \epsilon\ik\\ 
& + \sum_{k=1}^N\gamma^2 \eta^{(k)\top}\eta\ik + \sum_{k=1}^N\omega^2 v^\top v \\
\end{aligned}
\end{equation}
Integrating both sides of \eqref{eq:Vdot} on the interval $[0, T]$, we obtain
\begin{equation*}
\begin{aligned}
& V(\epsilon(T))+ \sum_{k=1}^N \int_0^T \epsilon^{(k)\top}W\ik \epsilon\ik dt \\
&\leq \sum_{k=1}^N \int_0^T \left( \omega^2 \| v \|^2 + \gamma^2 \| \eta\ik \|^2 \right) dt + \sum_{k=1}^N \epsilon^{(k)\top}_0 P\ik \epsilon\ik_0.
\end{aligned}
\end{equation*}
As $V(e(T))\geq 0$ and with the zero initial conditions of the observer states, it follows that
\begin{equation*}
 \sum_{k=1}^N \int_0^T \!\!\! \epsilon^{(k)\top}W\ik \epsilon\ik dt \leq  \sum_{k=1}^N \int_0^T \!\!\! \left( \omega^2 \| v \|^2 + \gamma^2 \| \eta\ik \|^2 \right) dt + I_0.
\end{equation*}
Letting $T\to \infty$, this satisfies Property (ii) of Problem 1.

Moreover, if $\xi_k=0$ and $\eta_k=0$ for all $k=1,...,N$, then it
follows from (\ref{eq:Vdot})
 that
\begin{equation*}
\dot{V}(\epsilon) \leq -\alpha V,
\end{equation*}
which implies that Property (i) of Problem 1 holds. 
}
\flushright $\blacksquare$

\end{Theorem}

Note that the choice of $\alpha$ determines the convergence speed of the
estimators, where a larger $\alpha$ enforces faster convergence of the
estimates.  However, larger values of $\alpha$ typically lead to higher
filter gain values. 

The salient feature of the resulting cooperative estimators
\eqref{eq:estimator} is that these estimators are
local and their complexity does not increase with the total size of the network. In this sense, the method presented in this paper is
scalable and guarantees $\mathcal{H}_\infty$-type performance. 
In contrast, a direct application of the algorithms developed in
\cite{OlfatiSaber:2005vm}, \cite{Olfati-saber2006} and
\cite{Ugrinovskii2011}, \cite{Ugrinovskii2011a}, to the problem
considered here would result in the order of the estimators growing with
the size of the network. 
Some remarks on the solution of the LMIs \eqref{LMI} are in order now.

\begin{Remark}
As it can be seen from the LMIs \eqref{LMI}, the solution to design problem
presented here involves solving coupled LMIs. When the nature of the
application allows for these LMis to be solved offline, this can be done in
a centralized manner. The resulting gain matrices $L\ik$, $K\ik$ can then
be deployed to the filters, this will ensure that while the estimation
algorithm is running, the estimators are fully distributed. Alternatively,
it was shown in \cite{Ugrinovskii2011a}, that such LMIs can be solved with
gradient descent type algorithms that allow distributed implementation.   
\end{Remark}


\begin{Remark}
As noted before, the choice of the partial state vectors
\eqref{sys:partition_xk} is not unique. As a special case, the choice $x\ik
= x$ for all $k=1,...,N$ yields local estimators similar to
\cite{Ugrinovskii2011,Ugrinovskii2011a}. 
\end{Remark}

 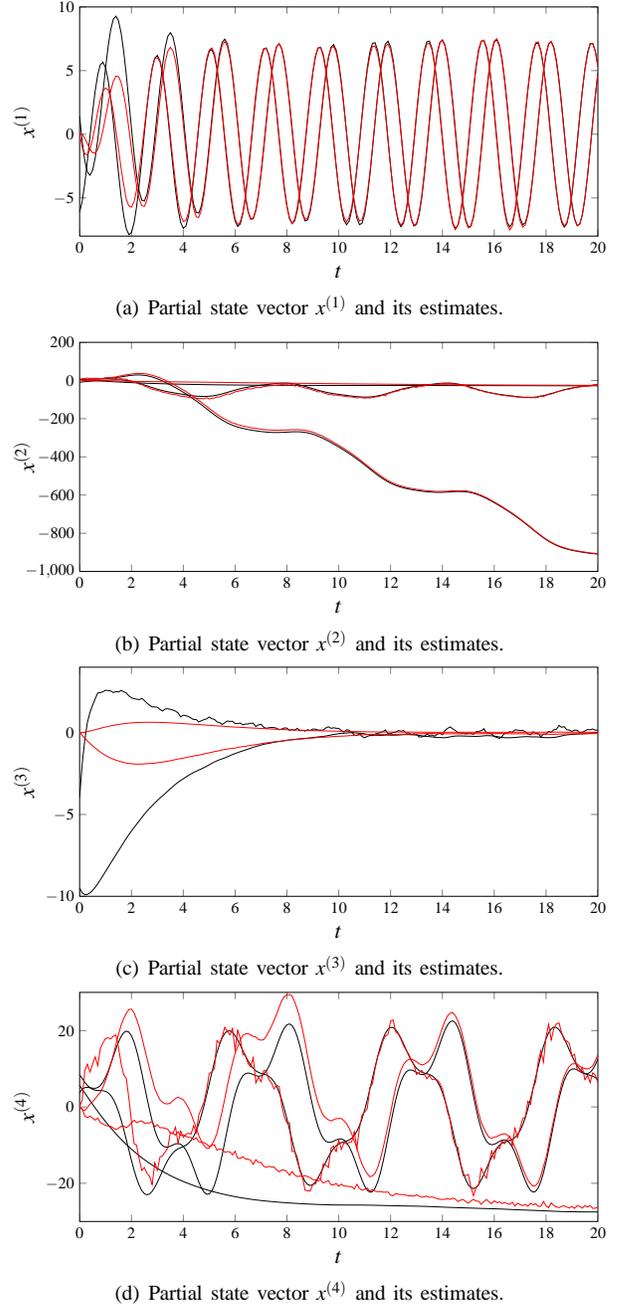
\begin{figure}
  \centering
 	\subfigure[Partial state vector $x^{(1)}$ and its estimates.]
 	{\input{x1.tex}}
 		\subfigure[Partial state vector $x^{(2)}$ and its estimates.]
 	{\input{x2.tex}}
 		\subfigure[Partial state vector $x^{(3)}$ and its estimates.]
 	{\input{x3.tex}}
 		\subfigure[Partial state vector $x^{(4)}$ and its estimates.]
 	{\input{x4.tex}}
 	\caption{Plots of the local states. Black lines represent the actual partial state vectors $x\ik$, red lines represent the estimated states $\hat{x}\ik$.}	
 	\label{fig:estimation}
 \end{figure}

\section{Simulation Example}

As an illustrative example, we consider a $9$-dimensional system, which is
composed of $4$ subsystems 

\begin{equation}\label{sys:example_dynamics}
\dot{x}=\!\!\left[\begin{array}{cc;{2pt/2pt}ccc;{2pt/2pt}cc;{2pt/2pt}cc}
0 & 3 & 0 & 0 & 0 & 0 & 1 & 0 & 0\\ \!\!-3 & 0 & 0 & 0 & 0 & \!\!-1 & 0 & 0 & 0\\ \hdashline[2pt/2pt] 
0 & 0 & 0 & 1 & 0 & 0 & 0 & 1 & 0 \\ 
0 & 0 & 0 & 0 & 0 & 0 & 0 & 0 & 2\\ 
0 & 0 & 0 & 0 & 0 & 1 & 0 & 0 & 0\\ \hdashline[2pt/2pt]
0 & 0 & 0 & 0 & 0 & 0 & 1 & 0 & 0\\ 
0 & 0 & 0 & 0 & 0 & \!\!-1 & -3 & 0 & 0\\ \hdashline[2pt/2pt] 
2 & 0 & 0 & 0 & 0 & 0 & 0 & 0 & 1\\ 
0 & 1 & 0 & 0 & 0 & 0 & 0 & \!\!-1 & 0 \end{array}\!\!\right] x
+\!\!\left[\begin{array}{cc}
0 & 0\\ 0 & 0\\ \hdashline[2pt/2pt] 0 & 0\\ 1 & 0 \\ 0 & 0\\ \hdashline[2pt/2pt] 0 & 0 \\ 0 & 1 \\ \hdashline[2pt/2pt] 0 & 0 \\ 0 & 0
\end{array}\!\!\right]v
\end{equation}
with $x=[x_{1,1} \; x_{1,2} \; x_{2,1} \; x_{2,2} \; x_{2,3} \;x_{3,1} \; x_{3,2} \;x_{4,1} \;x_{4,2} \; ]^\top$, and four measurements
\begin{equation}
\begin{aligned}\label{sys:example_output}
\begin{bmatrix} y_1 \\ y_2 \\ y_3 \\ y_4 \end{bmatrix} = \begin{bmatrix} 0 & 1 & 0 & 0 & 0 & 0 & 0 & 0 & 0\\ 0 & 0 & -1 & 0 & 1 & 0 & 0 & 0 & 0\\ 0 & 0 & 0 & 0 & 0 & 1 & 0 & 0 & 0\\ 0 & 0 & 0 & 0 & 1 & 0 & 0 & -1 & 0 \end{bmatrix} x + \eta,
\end{aligned}
\end{equation}
which are associated with the four subsystems.
Thus, the partial state vectors \eqref{sys:partition_xk} can be chosen as shown in Figure \ref{fig:estimator_setup}.
%
Note that although the system has a dimension of $9$ and all the coordinates of
the state vector are tightly coupled, none of the estimators need to handle
more than three coordinates. In particular, $(A^{(2)}, C^{(2)})$ as defined in \eqref{sys:global_k} is not
detectable. This emphasizes the benefits of our algorithm, which can deal
with situations in which subsystems are individually not detectable. 

The graphs representing the interconnection topology are shown in Figure \ref{fig:estimator_setup} and \ref{fig:extended_graph}. After applying the estimators \eqref{eq:estimator}, we obtain simulation results shown in Figure \ref{fig:estimation}. All local estimators obtain a correct estimation of their respective partial state vector. Moreover, property (ii) of Problem 1 is satisfied with performance values of $\gamma=11.1$ and $\omega=8.4$.

\section{CONCLUSION}
In this paper, we presented a $\mathcal{H}_\infty$-based approach to
cooperative state estimation for linear interconnected large-scale systems,
such as multi-agent systems. In order to achieve scalability of the
estimation setup, we required the local estimators to estimate local states
only. We establish an algorithm for interconnecting the local estimators,
whereby both physical couplings and detectability issues can be
handled. Moreover, design conditions are presented to guarantee
$\mathcal{H}_\infty$-performance with respect to both model and measurement
disturbances. 

Further research will include the distributed calculation of the filter
gains \eqref{eq:filtergains}.
Moreover, an
interesting problem to consider is concerned with the
application of the cooperative estimation algorithm to cooperative and decentralized control problems. This may
yield interesting results with respect to the $\mc{H}_\infty$-performance
that can still be guaranteed.

\bibliographystyle{unsrt}
\bibliography{bibliography}

\end{document}

%% file: x1.tex
%
%
%
%
\begin{tikzpicture}[scale=0.6]

\begin{axis}[%
width=4.52083333333333in,
height=2in,
scale only axis,
xmin=0,
xmax=20,
xlabel={\Large $t$},
ymin=-8,
ymax=10,
ylabel={\Large $x^{(1)}$}
]
\addplot [
color=black,
solid,
forget plot
]
table[row sep=crcr]{
0 1.41026676859035\\
0.1 -0.598155427209723\\
0.2 -2.08928235673091\\
0.3 -2.97483012223446\\
0.4 -3.21835028430896\\
0.5 -2.83131091923464\\
0.6 -1.86343493520386\\
0.7 -0.410449769165944\\
0.8 1.37160791521981\\
0.9 3.30253549059435\\
1 5.20601360667896\\
1.1 6.90039486002772\\
1.2 8.22121375045287\\
1.3 9.03842617734886\\
1.4 9.27694400517998\\
1.5 8.91083504697243\\
1.6 7.96698188989839\\
1.7 6.51927359762052\\
1.8 4.68162275098005\\
1.9 2.60552619015012\\
2 0.467568488406633\\
2.1 -1.54151800869915\\
2.2 -3.2441071850322\\
2.3 -4.49339515908131\\
2.4 -5.17804807015559\\
2.5 -5.24914376732807\\
2.6 -4.71754761657383\\
2.7 -3.63585127772584\\
2.8 -2.10325677992275\\
2.9 -0.254808494659096\\
3 1.74353976555352\\
3.1 3.70183302052247\\
3.2 5.43719413735724\\
3.3 6.80344519436031\\
3.4 7.67922616590516\\
3.5 7.97444095027572\\
3.6 7.65545517355116\\
3.7 6.74522016052779\\
3.8 5.31752680228357\\
3.9 3.48713711420369\\
4 1.42657229407094\\
4.1 -0.673900536952725\\
4.2 -2.6377663378049\\
4.3 -4.2955215859036\\
4.4 -5.50362694319562\\
4.5 -6.15197586352883\\
4.6 -6.18804276600153\\
4.7 -5.60642296246057\\
4.8 -4.45130297263037\\
4.9 -2.82914984346268\\
5 -0.884909286042977\\
5.1 1.19673211398827\\
5.2 3.22065922361884\\
5.3 5.00019230363323\\
5.4 6.37468479588796\\
5.5 7.22102553134188\\
5.6 7.47009694573425\\
5.7 7.10858680387329\\
5.8 6.1594876255117\\
5.9 4.70118773640833\\
6 2.85583665265106\\
6.1 0.787142965738013\\
6.2 -1.3141411742101\\
6.3 -3.26774193903886\\
6.4 -4.89878821202926\\
6.5 -6.0583243691627\\
6.6 -6.6440973186635\\
6.7 -6.60565938134359\\
6.8 -5.96096144717596\\
6.9 -4.76306439050384\\
7 -3.10906882000094\\
7.1 -1.16238020697195\\
7.2 0.89732461025316\\
7.3 2.89560474289006\\
7.4 4.65723045449414\\
7.5 6.02189182608518\\
7.6 6.86405323744798\\
7.7 7.1032872033911\\
7.8 6.72180693934504\\
7.9 5.7609351171307\\
8 4.30288567568193\\
8.1 2.46678452320793\\
8.2 0.419681126360742\\
8.3 -1.65176339874685\\
8.4 -3.55912873838549\\
8.5 -5.1395115586291\\
8.6 -6.25518879820292\\
8.7 -6.80246104514375\\
8.8 -6.72946023459536\\
8.9 -6.04378177980953\\
9 -4.80647188287249\\
9.1 -3.12684719235239\\
9.2 -1.16249036987537\\
9.3 0.910474570878337\\
9.4 2.90966721714197\\
9.5 4.66262779845574\\
9.6 6.01488836040428\\
9.7 6.83365205597478\\
9.8 7.04332202297329\\
9.9 6.6272413765254\\
10 5.62007095079025\\
10.1 4.10588271120025\\
10.2 2.21481966502127\\
10.3 0.129480736569014\\
10.4 -1.96144093765501\\
10.5 -3.88460984981469\\
10.6 -5.47535594751369\\
10.7 -6.58630585401415\\
10.8 -7.10340430621294\\
10.9 -6.98883606742708\\
11 -6.26100542751441\\
11.1 -4.97048884274068\\
11.2 -3.22612508201488\\
11.3 -1.18777439693444\\
11.4 0.965608371197624\\
11.5 3.04416792633286\\
11.6 4.85949318208492\\
11.7 6.25080767094833\\
11.8 7.09696060870001\\
11.9 7.30673038894255\\
12 6.85726087028785\\
12.1 5.80338047509835\\
12.2 4.24046758160224\\
12.3 2.30946429215667\\
12.4 0.178916702623373\\
12.5 -1.96367924827464\\
12.6 -3.93342151290595\\
12.7 -5.5550235598121\\
12.8 -6.68122277443074\\
12.9 -7.20956610464466\\
13 -7.08934285500989\\
13.1 -6.34071208135056\\
13.2 -5.02827157399037\\
13.3 -3.25893646482717\\
13.4 -1.18360896996277\\
13.5 1.00667757237755\\
13.6 3.11217915276456\\
13.7 4.94457969692307\\
13.8 6.34663762746687\\
13.9 7.19957765200385\\
14 7.41601427308683\\
14.1 6.98853749711431\\
14.2 5.96490870421014\\
14.3 4.42740225699432\\
14.4 2.49980639677455\\
14.5 0.344947155694499\\
14.6 -1.84566173067837\\
14.7 -3.87287697320076\\
14.8 -5.54962764143331\\
14.9 -6.72737520303099\\
15 -7.297201555126\\
15.1 -7.20685799324923\\
15.2 -6.47022215990405\\
15.3 -5.14379917122395\\
15.4 -3.34273504582141\\
15.5 -1.23323540960755\\
15.6 1.00126547344143\\
15.7 3.14480812016951\\
15.8 4.99959124513777\\
15.9 6.41012729668001\\
16 7.24366662983043\\
16.1 7.42580261797831\\
16.2 6.94594372072184\\
16.3 5.85676204082897\\
16.4 4.26940852193142\\
16.5 2.32002750402336\\
16.6 0.177385618501666\\
16.7 -1.97005176716863\\
16.8 -3.92632496772386\\
16.9 -5.51670540418918\\
17 -6.60849708671835\\
17.1 -7.10279860285312\\
17.2 -6.95832165745221\\
17.3 -6.19602888009593\\
17.4 -4.88219356545441\\
17.5 -3.11751169192333\\
17.6 -1.06324508280219\\
17.7 1.08869410927795\\
17.8 3.14084437389209\\
17.9 4.91985330232781\\
18 6.27570911045924\\
18.1 7.08879676567881\\
18.2 7.2832734554556\\
18.3 6.8416565089332\\
18.4 5.80591438581568\\
18.5 4.27312666832412\\
18.6 2.37108223553662\\
18.7 0.259951546204026\\
18.8 -1.87743989712254\\
18.9 -3.85309902856029\\
19 -5.48631632781896\\
19.1 -6.62556995735044\\
19.2 -7.1667299905153\\
19.3 -7.0728451845825\\
19.4 -6.35049758114477\\
19.5 -5.0538804392404\\
19.6 -3.29567030038097\\
19.7 -1.23822906400139\\
19.8 0.932804035132563\\
19.9 3.01917164401261\\
20 4.83281880783095\\
};
\addplot [
color=black,
solid,
forget plot
]
table[row sep=crcr]{
0 -6.1103758317668\\
0.1 -5.25511718838204\\
0.2 -3.85263281062429\\
0.3 -2.08724259009588\\
0.4 -0.158058198411929\\
0.5 1.73661150444866\\
0.6 3.40950290349507\\
0.7 4.69553046903038\\
0.8 5.46784407083112\\
0.9 5.65440304309171\\
1 5.23623436096542\\
1.1 4.24809534455765\\
1.2 2.77829064762049\\
1.3 0.959605360412354\\
1.4 -1.04317100376113\\
1.5 -3.05214199010437\\
1.6 -4.88810462096969\\
1.7 -6.38875633533025\\
1.8 -7.41809336864262\\
1.9 -7.88046480149587\\
2 -7.72855666418303\\
2.1 -6.97128038151242\\
2.2 -5.67396912380251\\
2.3 -3.95118291686171\\
2.4 -1.95635383794079\\
2.5 0.131359496267099\\
2.6 2.12974450798687\\
2.7 3.86615869537569\\
2.8 5.19063667446493\\
2.9 5.98836871652057\\
3 6.1888489953924\\
3.1 5.77559832844165\\
3.2 4.78983099048865\\
3.3 3.32212310170529\\
3.4 1.50089840286054\\
3.5 -0.512071531603631\\
3.6 -2.53582008407224\\
3.7 -4.38770178188737\\
3.8 -5.89954434873254\\
3.9 -6.93116286010259\\
4 -7.38410911753372\\
4.1 -7.21909383917583\\
4.2 -6.45031334189442\\
4.3 -5.14408926193412\\
4.4 -3.41372198183238\\
4.5 -1.41074091904288\\
4.6 0.687511580816446\\
4.7 2.69688187403648\\
4.8 4.43641402653142\\
4.9 5.74817552910255\\
5 6.51244518310166\\
5.1 6.65845139243877\\
5.2 6.17544976225267\\
5.3 5.10928624797556\\
5.4 3.55996383465061\\
5.5 1.66871076003757\\
5.6 -0.392921205823597\\
5.7 -2.44373966986074\\
5.8 -4.30380948660567\\
5.9 -5.80691127304573\\
6 -6.81783309776998\\
6.1 -7.24263281203471\\
6.2 -7.04286078946089\\
6.3 -6.23619200458424\\
6.4 -4.89246881016692\\
6.5 -3.13231877490744\\
6.6 -1.11370358245266\\
6.7 0.98179339573862\\
6.8 2.96805027555652\\
6.9 4.67231119650108\\
7 5.94070334186886\\
7.1 6.65956525867246\\
7.2 6.76968032389358\\
7.3 6.26390931727891\\
7.4 5.18687914875074\\
7.5 3.6341440798972\\
7.6 1.74476396701812\\
7.7 -0.310833693455103\\
7.8 -2.34660530270606\\
7.9 -4.1816975488302\\
8 -5.65418251426508\\
8.1 -6.63168021712712\\
8.2 -7.02353880575237\\
8.3 -6.79490046744259\\
8.4 -5.96632973880578\\
8.5 -4.61352779098642\\
8.6 -2.85420311786053\\
8.7 -0.845066612286575\\
8.8 1.23492849884229\\
8.9 3.19830800478049\\
9 4.86984895257497\\
9.1 6.09823747208894\\
9.2 6.7736877052543\\
9.3 6.83736224475233\\
9.4 6.28407940796954\\
9.5 5.16211505406909\\
9.6 3.56791730941683\\
9.7 1.64180963462998\\
9.8 -0.442278227462025\\
9.9 -2.49820300665129\\
10 -4.34168569324367\\
10.1 -5.8068059419984\\
10.2 -6.75858084297643\\
10.3 -7.10947414687811\\
10.4 -6.83109491434232\\
10.5 -5.94756039305879\\
10.6 -4.53318147573505\\
10.7 -2.70885118566489\\
10.8 -0.636577467631093\\
10.9 1.49686415004968\\
11 3.50552634120086\\
11.1 5.21245617068337\\
11.2 6.46292962814015\\
11.3 7.14343384176152\\
11.4 7.19230719039688\\
11.5 6.60240119133633\\
11.6 5.42437889383407\\
11.7 3.76181509716548\\
11.8 1.76046256531942\\
11.9 -0.402351614061358\\
12 -2.52812097382422\\
12.1 -4.42592154807811\\
12.2 -5.92837443177089\\
12.3 -6.90302645140231\\
12.4 -7.26525872617178\\
12.5 -6.98247463121211\\
12.6 -6.07956673233419\\
12.7 -4.63342945809124\\
12.8 -2.77187259913386\\
12.9 -0.659000295865799\\
13 1.51633998634032\\
13.1 3.56129579085015\\
13.2 5.29754607646908\\
13.3 6.57069382691625\\
13.4 7.26507817907724\\
13.5 7.31556342371955\\
13.6 6.71855929167026\\
13.7 5.52776619613659\\
13.8 3.8504961226265\\
13.9 1.83325770038636\\
14 -0.345407654135362\\
14.1 -2.49057970027858\\
14.2 -4.41801482007134\\
14.3 -5.96105070573825\\
14.4 -6.98410348565046\\
14.5 -7.39238309365349\\
14.6 -7.14525391193258\\
14.7 -6.26058500517797\\
14.8 -4.81585614415916\\
14.9 -2.93944313476043\\
15 -0.798017694119672\\
15.1 1.41602341388628\\
15.2 3.50543546635314\\
15.3 5.28423021712991\\
15.4 6.58968199539754\\
15.5 7.30337049586232\\
15.6 7.35939706649633\\
15.7 6.75100846070805\\
15.8 5.5386801760392\\
15.9 3.83323319640936\\
16 1.7893179289044\\
16.1 -0.405600219894721\\
16.2 -2.55253867333188\\
16.3 -4.45866225875563\\
16.4 -5.95781896859262\\
16.5 -6.92234569420934\\
16.6 -7.2681224153116\\
16.7 -6.96623873882063\\
16.8 -6.04417381702458\\
16.9 -4.58799060607396\\
17 -2.72800212851583\\
17.1 -0.629489372460223\\
17.2 1.52044542066339\\
17.3 3.53268933432153\\
17.4 5.23208800169941\\
17.5 6.46769856066662\\
17.6 7.12538083880943\\
17.7 7.1479290253721\\
17.8 6.53552395993993\\
17.9 5.34690610700546\\
18 3.68527709590186\\
18.1 1.69687457634584\\
18.2 -0.44475928629407\\
18.3 -2.54920043416156\\
18.4 -4.43227116644078\\
18.5 -5.92907487945096\\
18.6 -6.91149045816589\\
18.7 -7.29056967756747\\
18.8 -7.03141224639936\\
18.9 -6.15228318606057\\
19 -4.72931841873781\\
19.1 -2.88794837925949\\
19.2 -0.794296914651874\\
19.3 1.36539750393798\\
19.4 3.40176452963068\\
19.5 5.13277964155959\\
19.6 6.40103819862364\\
19.7 7.090616298373\\
19.8 7.13940817954884\\
19.9 6.54237656804489\\
20 5.3545340765377\\
};
\addplot [
color=red,
solid,
forget plot
]
table[row sep=crcr]{
0 0\\
0.1 -0.000612107410335044\\
0.2 -0.30055127433424\\
0.3 -0.741093132150778\\
0.4 -1.17583573634801\\
0.5 -1.45545719903309\\
0.6 -1.48596261143799\\
0.7 -1.19561925043926\\
0.8 -0.569295623718724\\
0.9 0.304907966665272\\
1 1.34478732052341\\
1.1 2.41611147070993\\
1.2 3.40596205396996\\
1.3 4.16010990676852\\
1.4 4.57254047099776\\
1.5 4.5324853392809\\
1.6 4.07265930189246\\
1.7 3.18057667857789\\
1.8 1.95459376536189\\
1.9 0.446262679262718\\
2 -1.14973885531397\\
2.1 -2.7044847862337\\
2.2 -4.06312881868706\\
2.3 -5.07980314590627\\
2.4 -5.62429352515487\\
2.5 -5.65522999863756\\
2.6 -5.17102970898802\\
2.7 -4.18482000887694\\
2.8 -2.76245849134343\\
2.9 -1.02478408986076\\
3 0.865240598066412\\
3.1 2.69437356279855\\
3.2 4.35007537952855\\
3.3 5.66934954866569\\
3.4 6.51586743986404\\
3.5 6.82482567944631\\
3.6 6.55611859778766\\
3.7 5.73998867801583\\
3.8 4.4404169946969\\
3.9 2.73362235126893\\
4 0.80720435177186\\
4.1 -1.18019723815894\\
4.2 -3.04863226849651\\
4.3 -4.66044801300206\\
4.4 -5.84893691527313\\
4.5 -6.49659605937852\\
4.6 -6.52388576798984\\
4.7 -5.92223178986966\\
4.8 -4.75716810939184\\
4.9 -3.13353391712928\\
5 -1.18672489381315\\
5.1 0.883016795062391\\
5.2 2.90769991407922\\
5.3 4.70990063887313\\
5.4 6.11580784927346\\
5.5 7.00719818221748\\
5.6 7.27516902474574\\
5.7 6.92649470184167\\
5.8 6.02486341220297\\
5.9 4.60892604690572\\
6 2.81020978990229\\
6.1 0.7596982217586\\
6.2 -1.32964958509592\\
6.3 -3.26026021805734\\
6.4 -4.88583073111476\\
6.5 -6.06292720487761\\
6.6 -6.67646017767843\\
6.7 -6.6519327052901\\
6.8 -5.98456996078181\\
6.9 -4.75568209466933\\
7 -3.08279809699478\\
7.1 -1.12027871205196\\
7.2 0.952006854857901\\
7.3 2.94215644974198\\
7.4 4.68285714514915\\
7.5 6.04190744309864\\
7.6 6.88212452265498\\
7.7 7.12636716142315\\
7.8 6.72757767203048\\
7.9 5.71496063925522\\
8 4.20693461787112\\
8.1 2.37240192172277\\
8.2 0.345255599340751\\
8.3 -1.71688243885738\\
8.4 -3.59539899491786\\
8.5 -5.12005341610877\\
8.6 -6.1795331786536\\
8.7 -6.68463703825981\\
8.8 -6.58167128447076\\
8.9 -5.88114203205915\\
9 -4.64489404300852\\
9.1 -2.98707249460149\\
9.2 -1.02067947818214\\
9.3 1.04189768971526\\
9.4 3.0200473814536\\
9.5 4.71947996764384\\
9.6 5.99189809542296\\
9.7 6.74507402214681\\
9.8 6.90024584766762\\
9.9 6.45455830476826\\
10 5.43888281527964\\
10.1 3.94254864848706\\
10.2 2.10458761955087\\
10.3 0.103786680052269\\
10.4 -1.92698473106331\\
10.5 -3.76007509214719\\
10.6 -5.27616059322132\\
10.7 -6.32396270937115\\
10.8 -6.81789227384407\\
10.9 -6.71683167219443\\
11 -5.99337650324345\\
11.1 -4.72709082712683\\
11.2 -3.03508399979686\\
11.3 -1.07038523026636\\
11.4 1.02090242303366\\
11.5 3.0264030927041\\
11.6 4.77045289454562\\
11.7 6.10523624106541\\
11.8 6.88889193059224\\
11.9 7.08300877534501\\
12 6.64020931598894\\
12.1 5.59409360803295\\
12.2 4.03691061716596\\
12.3 2.14522050188514\\
12.4 0.0492231099320752\\
12.5 -2.05868884547811\\
12.6 -3.98981236743773\\
12.7 -5.56276375873087\\
12.8 -6.64613943482831\\
12.9 -7.14994336951117\\
13 -7.00710707976431\\
13.1 -6.21455874808455\\
13.2 -4.86854458722619\\
13.3 -3.09127865366733\\
13.4 -1.03420884659592\\
13.5 1.10973133060197\\
13.6 3.17027108410649\\
13.7 4.99014247198589\\
13.8 6.37590381126549\\
13.9 7.18601363284679\\
14 7.37071814483602\\
14.1 6.92313524671775\\
14.2 5.86755505017234\\
14.3 4.29921464914919\\
14.4 2.34950941877085\\
14.5 0.1643576215097\\
14.6 -2.05645218222944\\
14.7 -4.0677112900368\\
14.8 -5.71354661872522\\
14.9 -6.82762700414747\\
15 -7.36362642032872\\
15.1 -7.25372653121558\\
15.2 -6.49032887295261\\
15.3 -5.15640984922441\\
15.4 -3.33732622929894\\
15.5 -1.21233840944344\\
15.6 1.00625381955853\\
15.7 3.15098629697132\\
15.8 5.03533685964957\\
15.9 6.47638479923538\\
16 7.31126398735062\\
16.1 7.50564431013299\\
16.2 7.03203143590727\\
16.3 5.91732935430212\\
16.4 4.29253276934084\\
16.5 2.28864861370949\\
16.6 0.0915023163232147\\
16.7 -2.12934152244569\\
16.8 -4.16176182663279\\
16.9 -5.77653518857947\\
17 -6.82699616824926\\
17.1 -7.29039635455914\\
17.2 -7.11882809986969\\
17.3 -6.31199752929226\\
17.4 -4.97687289015921\\
17.5 -3.1857661785434\\
17.6 -1.11009415890416\\
17.7 1.06916971642493\\
17.8 3.17248765355617\\
17.9 4.96821333498901\\
18 6.32367520104529\\
18.1 7.11647315676166\\
18.2 7.26785797783698\\
18.3 6.76050411668073\\
18.4 5.68074610639886\\
18.5 4.13773229458922\\
18.6 2.20700917977693\\
18.7 0.11558987821296\\
18.8 -2.01047887640391\\
18.9 -3.97073728822618\\
19 -5.58700465263228\\
19.1 -6.68661218096243\\
19.2 -7.19084414307524\\
19.3 -7.04998795750832\\
19.4 -6.28714775208656\\
19.5 -4.99085694038202\\
19.6 -3.26120933663842\\
19.7 -1.25912421346101\\
19.8 0.908278389726822\\
19.9 2.97401450059976\\
20 4.77061504160308\\
};
\addplot [
color=red,
solid,
forget plot
]
table[row sep=crcr]{
0 0\\
0.1 -1.01495940288099\\
0.2 -1.54651015857092\\
0.3 -1.61000306251592\\
0.4 -1.17252475611987\\
0.5 -0.395818574089184\\
0.6 0.659607776446071\\
0.7 1.81248486326095\\
0.8 2.71979143048518\\
0.9 3.38730615661554\\
1 3.62858014720197\\
1.1 3.4939535294237\\
1.2 2.82888854919507\\
1.3 1.77122101222471\\
1.4 0.291864621176833\\
1.5 -1.14536973458045\\
1.6 -2.68451637303654\\
1.7 -3.9571024000882\\
1.8 -5.10122940375559\\
1.9 -5.63016468419275\\
2 -5.7339449201658\\
2.1 -5.30723221132623\\
2.2 -4.35632393670274\\
2.3 -2.9137425966315\\
2.4 -1.25922807104443\\
2.5 0.479582561732583\\
2.6 2.25178788574336\\
2.7 3.87688465220391\\
2.8 5.1587917895207\\
2.9 5.93648476455039\\
3 6.01913767654285\\
3.1 5.7103384737061\\
3.2 4.82372026140465\\
3.3 3.42667730708098\\
3.4 1.73896854084595\\
3.5 -0.165101332506426\\
3.6 -2.053457933053\\
3.7 -3.8082411718918\\
3.8 -5.37624133393244\\
3.9 -6.37289460471196\\
4 -6.86480883836335\\
4.1 -6.75790242560653\\
4.2 -6.16592874663225\\
4.3 -4.97187014231087\\
4.4 -3.31544906841226\\
4.5 -1.30906713717494\\
4.6 0.81941619485979\\
4.7 2.82186773462972\\
4.8 4.555354375324\\
4.9 5.89902018381908\\
5 6.61500084240011\\
5.1 6.78087069764114\\
5.2 6.33931043236098\\
5.3 5.27618552903859\\
5.4 3.75241403559402\\
5.5 1.78455225227099\\
5.6 -0.255463059963231\\
5.7 -2.17564317077576\\
5.8 -4.04756625971809\\
5.9 -5.55371431169482\\
6 -6.67278947715642\\
6.1 -7.10881253481248\\
6.2 -6.88341242639603\\
6.3 -6.1391035281476\\
6.4 -4.86300815009508\\
6.5 -3.12923095943101\\
6.6 -1.05865081024987\\
6.7 1.12376484581984\\
6.8 3.12852055522502\\
6.9 4.82084497346846\\
7 6.05810800718168\\
7.1 6.73014144620185\\
7.2 6.76983591209137\\
7.3 6.23085574914967\\
7.4 5.20586336787679\\
7.5 3.65899710447621\\
7.6 1.77209417447884\\
7.7 -0.364312634437723\\
7.8 -2.49803243656508\\
7.9 -4.32331328473291\\
8 -5.65373620194921\\
8.1 -6.58048646206204\\
8.2 -7.00165905431867\\
8.3 -6.68939337406664\\
8.4 -5.77395326472108\\
8.5 -4.42777705318852\\
8.6 -2.7092841139958\\
8.7 -0.725080509147671\\
8.8 1.3151846303786\\
8.9 3.23611275166803\\
9 4.85256637783271\\
9.1 6.14865667167089\\
9.2 6.77348966921938\\
9.3 6.80092165424161\\
9.4 6.15827294511968\\
9.5 4.97208558593837\\
9.6 3.4080087461453\\
9.7 1.50046602094906\\
9.8 -0.507746243675868\\
9.9 -2.50263833115017\\
10 -4.28113073706681\\
10.1 -5.66153722911962\\
10.2 -6.48987010539078\\
10.3 -6.89057405136415\\
10.4 -6.52767981599794\\
10.5 -5.73002061806067\\
10.6 -4.3731999078641\\
10.7 -2.65504241847222\\
10.8 -0.713197537041577\\
10.9 1.40948483618128\\
11 3.35734304913447\\
11.1 4.990522763531\\
11.2 6.16995773015885\\
11.3 6.89455664702984\\
11.4 6.92091232307415\\
11.5 6.34006722688871\\
11.6 5.2182557208793\\
11.7 3.55578183445463\\
11.8 1.67803928956724\\
11.9 -0.447991694952314\\
12 -2.55613608017704\\
12.1 -4.44249359868022\\
12.2 -5.81034931599288\\
12.3 -6.78692329536344\\
12.4 -7.14349608070674\\
12.5 -6.86285631993115\\
12.6 -5.94215651483148\\
12.7 -4.52250608020435\\
12.8 -2.72381187406653\\
12.9 -0.610747931923596\\
13 1.60995864795129\\
13.1 3.60387294025769\\
13.2 5.26754468818036\\
13.3 6.48362237199115\\
13.4 7.08984513409547\\
13.5 7.13299053507418\\
13.6 6.62905567351841\\
13.7 5.43407294962261\\
13.8 3.69586291592668\\
13.9 1.70171706774151\\
14 -0.41103278078765\\
14.1 -2.5363754936781\\
14.2 -4.43151927550495\\
14.3 -5.9602468240593\\
14.4 -7.05154454421714\\
14.5 -7.50135171170007\\
14.6 -7.12890238532645\\
14.7 -6.19582591579173\\
14.8 -4.64627932072735\\
14.9 -2.85785443593168\\
15 -0.752770323653105\\
15.1 1.47497525546412\\
15.2 3.52254008988171\\
15.3 5.36568419137279\\
15.4 6.67528753034248\\
15.5 7.31425520534374\\
15.6 7.39972152757423\\
15.7 6.84245436127954\\
15.8 5.62651555746205\\
15.9 3.79456521903305\\
16 1.75355461092164\\
16.1 -0.47759406727099\\
16.2 -2.71741266694275\\
16.3 -4.61288125185955\\
16.4 -6.14187096375251\\
16.5 -7.09988560183945\\
16.6 -7.51243115473378\\
16.7 -7.20998154167776\\
16.8 -6.10400556774665\\
16.9 -4.44043539403859\\
17 -2.61236149147339\\
17.1 -0.533337290728954\\
17.2 1.64123541961354\\
17.3 3.55172656781074\\
17.4 5.29593709791969\\
17.5 6.52982566839373\\
17.6 7.19476399377444\\
17.7 7.26576151323655\\
17.8 6.5430754772247\\
17.9 5.32411616246074\\
18 3.62415605850751\\
18.1 1.57272363181173\\
18.2 -0.633278158740239\\
18.3 -2.64702040872756\\
18.4 -4.37928729163214\\
18.5 -5.92270917504789\\
18.6 -6.75954295987228\\
18.7 -7.19165653836695\\
18.8 -6.95111783698208\\
18.9 -6.08222567062469\\
19 -4.57828253679814\\
19.1 -2.72923516310819\\
19.2 -0.622457393933069\\
19.3 1.5034445907939\\
19.4 3.42466138618167\\
19.5 5.08700342859475\\
19.6 6.27252222938673\\
19.7 7.13268074144417\\
19.8 7.11708336607721\\
19.9 6.5188332390867\\
20 5.32878059850709\\
};
\end{axis}
\end{tikzpicture}%

%% file: x2.tex
%
%
%
%
\begin{tikzpicture}[scale=0.6]

\begin{axis}[%
width=4.52083333333333in,
height=2in,
scale only axis,
xmin=0,
xmax=20,
xlabel={\Large $t$},
ymin=-1000,
ymax=200,
ylabel={\Large $x^{(2)}$}
]
\addplot [
color=black,
solid,
forget plot
]
table[row sep=crcr]{
0 7.34078421222469\\
0.1 6.91443372613015\\
0.2 6.71310375567163\\
0.3 6.67286736596762\\
0.4 6.74638437027639\\
0.5 6.91540830017185\\
0.6 7.16669003764262\\
0.7 7.50690452188242\\
0.8 7.96472069080019\\
0.9 8.58221425357604\\
1 9.39684740489246\\
1.1 10.4410296157945\\
1.2 11.73814400814\\
1.3 13.2908005588699\\
1.4 15.0879540359816\\
1.5 17.0897119219729\\
1.6 19.2175132339469\\
1.7 21.3823839512324\\
1.8 23.501467933933\\
1.9 25.4641872263075\\
2 27.1333367061629\\
2.1 28.3836762526026\\
2.2 29.1008932110262\\
2.3 29.1863812744025\\
2.4 28.562072991263\\
2.5 27.175522988944\\
2.6 25.0107439014271\\
2.7 22.07126790101\\
2.8 18.3843103553402\\
2.9 14.004125715287\\
3 8.99405878398178\\
3.1 3.42628171214604\\
3.2 -2.62047027696888\\
3.3 -9.06494975602071\\
3.4 -15.8408887262956\\
3.5 -22.8802389399753\\
3.6 -30.1332258007798\\
3.7 -37.5945194963517\\
3.8 -45.2729612703488\\
3.9 -53.1808206782937\\
4 -61.3359631753085\\
4.1 -69.7711789325037\\
4.2 -78.5278719255195\\
4.3 -87.6329404567994\\
4.4 -97.090181612176\\
4.5 -106.880673331279\\
4.6 -116.973744904411\\
4.7 -127.31601810728\\
4.8 -137.831897311917\\
4.9 -148.427232679155\\
5 -158.98346575886\\
5.1 -169.383205723687\\
5.2 -179.510249561567\\
5.3 -189.231188963465\\
5.4 -198.424803302279\\
5.5 -206.995805998593\\
5.6 -214.876603499822\\
5.7 -222.035320431118\\
5.8 -228.463103177707\\
5.9 -234.171164097693\\
6 -239.195532871572\\
6.1 -243.596439437182\\
6.2 -247.441733153694\\
6.3 -250.814523372009\\
6.4 -253.814640310437\\
6.5 -256.518819208363\\
6.6 -258.97246846719\\
6.7 -261.225703235985\\
6.8 -263.311447298788\\
6.9 -265.238459308931\\
7 -266.993798222109\\
7.1 -268.547877764913\\
7.2 -269.877845780137\\
7.3 -270.944355576414\\
7.4 -271.699697454317\\
7.5 -272.125364355136\\
7.6 -272.23159963729\\
7.7 -272.025949949368\\
7.8 -271.55419963511\\
7.9 -270.893331864849\\
8 -270.125087382602\\
8.1 -269.358237772441\\
8.2 -268.715963203148\\
8.3 -268.334495313224\\
8.4 -268.334863079668\\
8.5 -268.823905973069\\
8.6 -269.910712486168\\
8.7 -271.659203677754\\
8.8 -274.105205027482\\
8.9 -277.28499962915\\
9 -281.206352376874\\
9.1 -285.840645199229\\
9.2 -291.129742031645\\
9.3 -297.000399534697\\
9.4 -303.375659451885\\
9.5 -310.177754790432\\
9.6 -317.327150389554\\
9.7 -324.756579721721\\
9.8 -332.405842556017\\
9.9 -340.241723583772\\
10 -348.253153088034\\
10.1 -356.429492163599\\
10.2 -364.798733693751\\
10.3 -373.399956671933\\
10.4 -382.266725694499\\
10.5 -391.435625093237\\
10.6 -400.934473739424\\
10.7 -410.762506480789\\
10.8 -420.901754366079\\
10.9 -431.327645889822\\
11 -441.982114441223\\
11.1 -452.78332431004\\
11.2 -463.634008875606\\
11.3 -474.415617417135\\
11.4 -484.989952816221\\
11.5 -495.216287696286\\
11.6 -504.976078455729\\
11.7 -514.182229354825\\
11.8 -522.738318151853\\
11.9 -530.568593907765\\
12 -537.637948539102\\
12.1 -543.935369492239\\
12.2 -549.491062296538\\
12.3 -554.36155211858\\
12.4 -558.606437308366\\
12.5 -562.30592730379\\
12.6 -565.556232108118\\
12.7 -568.433012339167\\
12.8 -571.007603293058\\
12.9 -573.355315951323\\
13 -575.523280896071\\
13.1 -577.528342875615\\
13.2 -579.374385666789\\
13.3 -581.045234435529\\
13.4 -582.519734273443\\
13.5 -583.766123326267\\
13.6 -584.747159804736\\
13.7 -585.425307426752\\
13.8 -585.768400765973\\
13.9 -585.780039948742\\
14 -585.485365778504\\
14.1 -584.925722733767\\
14.2 -584.171121988193\\
14.3 -583.315240120143\\
14.4 -582.473379968256\\
14.5 -581.771409245503\\
14.6 -581.345905947082\\
14.7 -581.326076916611\\
14.8 -581.819336023504\\
14.9 -582.931818684946\\
15 -584.749061018969\\
15.1 -587.312110331457\\
15.2 -590.643650674386\\
15.3 -594.744267383689\\
15.4 -599.573218971473\\
15.5 -605.07642062879\\
15.6 -611.179690421247\\
15.7 -617.786326358003\\
15.8 -624.806918558073\\
15.9 -632.162934157554\\
16 -639.793139805764\\
16.1 -647.649668387688\\
16.2 -655.700133333986\\
16.3 -663.930132765252\\
16.4 -672.333287984711\\
16.5 -680.926052388745\\
16.6 -689.746659917148\\
16.7 -698.818924201621\\
16.8 -708.174804541039\\
16.9 -717.850590634268\\
17 -727.852349800537\\
17.1 -738.168271455902\\
17.2 -748.759206369304\\
17.3 -759.577964162837\\
17.4 -770.553712484276\\
17.5 -781.580323559482\\
17.6 -792.529366015796\\
17.7 -803.264686649717\\
17.8 -813.659285386981\\
17.9 -823.590413609642\\
18 -832.943990063991\\
18.1 -841.634172679115\\
18.2 -849.612022947605\\
18.3 -856.846656307915\\
18.4 -863.330774403518\\
18.5 -869.089017555881\\
18.6 -874.163067119099\\
18.7 -878.62146168218\\
18.8 -882.546721283732\\
18.9 -886.011218199424\\
19 -889.105706695533\\
19.1 -891.91668689576\\
19.2 -894.50676448927\\
19.3 -896.925549425279\\
19.4 -899.186768260994\\
19.5 -901.293858778167\\
19.6 -903.247885673444\\
19.7 -905.025117742328\\
19.8 -906.588746078902\\
19.9 -907.895908857751\\
20 -908.916457560966\\
};
\addplot [
color=black,
solid,
forget plot
]
table[row sep=crcr]{
0 -9.33134954902366\\
0.1 -7.71410456525225\\
0.2 -6.08993884187899\\
0.3 -4.96702982330917\\
0.4 -3.77370356248451\\
0.5 -2.79319062130829\\
0.6 -1.93494886727158\\
0.7 -1.16018787140395\\
0.8 -0.288583470508061\\
0.9 0.622851410910804\\
1 1.48351830507716\\
1.1 2.31816582674432\\
1.2 3.04782690695733\\
1.3 3.5442573629389\\
1.4 3.93451878002084\\
1.5 3.82342017923416\\
1.6 3.15181456587081\\
1.7 2.07233737531119\\
1.8 0.813792980251292\\
1.9 -1.21231416888524\\
2 -3.83871471355914\\
2.1 -6.97808914216979\\
2.2 -10.585647535951\\
2.3 -14.5736766186045\\
2.4 -18.8803794020271\\
2.5 -23.3756996889217\\
2.6 -27.8099672994581\\
2.7 -32.3900478800607\\
2.8 -36.7566726286828\\
2.9 -40.9986494975319\\
3 -45.044793446418\\
3.1 -48.9021297452396\\
3.2 -52.4469022480797\\
3.3 -55.6721628449172\\
3.4 -58.7244323585851\\
3.5 -61.1964699775957\\
3.6 -63.5890154820004\\
3.7 -65.9993741171117\\
3.8 -68.332990282908\\
3.9 -70.5090658862239\\
4 -72.5360468396629\\
4.1 -74.6047203105162\\
4.2 -76.6923048131077\\
4.3 -78.6180934918616\\
4.4 -80.2729985420372\\
4.5 -81.5560040093544\\
4.6 -82.630385449306\\
4.7 -83.200792051421\\
4.8 -83.4388274283581\\
4.9 -83.0939646146923\\
5 -82.1621433070233\\
5.1 -80.824369666312\\
5.2 -78.9787243215312\\
5.3 -76.3340539473705\\
5.4 -73.2763172030422\\
5.5 -69.7023726684381\\
5.6 -65.9665694850537\\
5.7 -62.0866696150746\\
5.8 -58.1522348536311\\
5.9 -54.1340223805813\\
6 -50.2567386811925\\
6.1 -46.5429182446372\\
6.2 -42.9705711711065\\
6.3 -39.8129780288683\\
6.4 -37.1732675278415\\
6.5 -34.619900918313\\
6.6 -32.1693248060461\\
6.7 -30.1654381517519\\
6.8 -28.2142930141244\\
6.9 -26.5155432312456\\
7 -24.6976432039344\\
7.1 -23.0106008718739\\
7.2 -21.4707009491318\\
7.3 -19.7730165684187\\
7.4 -18.0062759289424\\
7.5 -16.5279525774176\\
7.6 -15.3377029930095\\
7.7 -14.1021889823051\\
7.8 -13.6021406566082\\
7.9 -13.3569409979946\\
8 -13.6421807061221\\
8.1 -14.469689737621\\
8.2 -15.9303980806567\\
8.3 -18.0986906595454\\
8.4 -20.6006803197286\\
8.5 -23.6316170794\\
8.6 -27.235838853828\\
8.7 -30.7352211959065\\
8.8 -34.5712922068884\\
8.9 -38.6655319240318\\
9 -42.8780270887719\\
9.1 -46.9422745513391\\
9.2 -50.7990772612518\\
9.3 -54.3898694230218\\
9.4 -57.7914750341304\\
9.5 -60.9077069628899\\
9.6 -63.7045920204493\\
9.7 -66.2845729719231\\
9.8 -68.4414359636278\\
9.9 -70.6351217217433\\
10 -72.5516652294578\\
10.1 -74.2331173992749\\
10.2 -76.1720894301859\\
10.3 -77.9264310749783\\
10.4 -79.7466353077913\\
10.5 -81.4565057588104\\
10.6 -83.1606730218539\\
10.7 -84.370446858162\\
10.8 -85.4908937374856\\
10.9 -86.2984788649222\\
11 -86.6785510206669\\
11.1 -86.59059421313\\
11.2 -86.0538864583304\\
11.3 -84.8495901220576\\
11.4 -82.9438584342518\\
11.5 -80.3626570516582\\
11.6 -77.4636128796519\\
11.7 -74.3572035177717\\
11.8 -70.4985667361012\\
11.9 -66.5312617230073\\
12 -62.263083605336\\
12.1 -57.993514760393\\
12.2 -53.8970703948799\\
12.3 -50.0242803621778\\
12.4 -46.1572054435001\\
12.5 -42.7855491773593\\
12.6 -39.6972085171683\\
12.7 -36.7490192023209\\
12.8 -34.1455183725007\\
12.9 -31.9871635383125\\
13 -29.8900160488824\\
13.1 -27.9396602374739\\
13.2 -26.0920813423063\\
13.3 -24.2588977219286\\
13.4 -22.6426278572863\\
13.5 -20.964938098543\\
13.6 -19.4352180403026\\
13.7 -17.7780727742545\\
13.8 -16.2212376200748\\
13.9 -15.0173332624681\\
14 -14.0496744973875\\
14.1 -13.4248604192962\\
14.2 -13.2727471969712\\
14.3 -13.626847918957\\
14.4 -14.6243817560173\\
14.5 -16.1478678598811\\
14.6 -18.4271878933733\\
14.7 -21.0865974149268\\
14.8 -24.1741638388948\\
14.9 -27.8239946733337\\
15 -31.7785543173793\\
15.1 -35.7526548028743\\
15.2 -40.0402958254943\\
15.3 -44.2774870859448\\
15.4 -48.343534543293\\
15.5 -52.3744752968316\\
15.6 -56.0295427073075\\
15.7 -59.2989509636153\\
15.8 -62.3283865770547\\
15.9 -65.0767706249796\\
16 -67.7337951856472\\
16.1 -70.1108246512003\\
16.2 -72.3993687598641\\
16.3 -74.4528275855207\\
16.4 -76.2876683861937\\
16.5 -78.1066871721886\\
16.6 -79.975308609867\\
16.7 -81.458363286837\\
16.8 -83.1686599969903\\
16.9 -84.6929591327098\\
17 -86.040894177014\\
17.1 -87.1138091285549\\
17.2 -87.7795750694386\\
17.3 -88.3438134623815\\
17.4 -88.349322854611\\
17.5 -87.8303543700617\\
17.6 -86.5300682535768\\
17.7 -84.6877113268056\\
17.8 -82.2815559732739\\
17.9 -79.3476409376313\\
18 -75.8571117239376\\
18.1 -72.1551566369206\\
18.2 -68.3210799808853\\
18.3 -64.2820848173894\\
18.4 -60.2170773985807\\
18.5 -56.2405315729067\\
18.6 -52.2818124039275\\
18.7 -48.719453364058\\
18.8 -45.3098553266074\\
18.9 -42.0452694692147\\
19 -39.3516896385345\\
19.1 -36.8631136708651\\
19.2 -34.6920156721322\\
19.3 -32.7455327466593\\
19.4 -30.7016915323997\\
19.5 -28.9775981864982\\
19.6 -27.381687834092\\
19.7 -25.8257430636654\\
19.8 -24.271481580887\\
19.9 -22.6770366458211\\
20 -21.2796651830945\\
};
\addplot [
color=black,
solid,
forget plot
]
table[row sep=crcr]{
0 5.60326300095945\\
0.1 4.63771027544997\\
0.2 3.65256276478974\\
0.3 2.66269113557899\\
0.4 1.67852727960912\\
0.5 0.706614155234748\\
0.6 -0.248510442190662\\
0.7 -1.18290422172118\\
0.8 -2.09354977789726\\
0.9 -2.9796801798533\\
1 -3.84071395513887\\
1.1 -4.67599642653087\\
1.2 -5.48553739323113\\
1.3 -6.26971227551187\\
1.4 -7.02910966325068\\
1.5 -7.76347490893413\\
1.6 -8.47274556618321\\
1.7 -9.1564923432752\\
1.8 -9.81520848486652\\
1.9 -10.4498262557182\\
2 -11.0618606909934\\
2.1 -11.6524851727955\\
2.2 -12.222281849465\\
2.3 -12.7715843239338\\
2.4 -13.3005088928697\\
2.5 -13.8088104167697\\
2.6 -14.2975670962079\\
2.7 -14.7682569919963\\
2.8 -15.222175660858\\
2.9 -15.6602055748912\\
3 -16.0825358627565\\
3.1 -16.4895259949365\\
3.2 -16.8822597565041\\
3.3 -17.2613729647734\\
3.4 -17.6261884275149\\
3.5 -17.9764764073314\\
3.6 -18.3125277711169\\
3.7 -18.6348135158881\\
3.8 -18.944025926814\\
3.9 -19.2414584061966\\
4 -19.5286994336934\\
4.1 -19.8054516333897\\
4.2 -20.0718525562149\\
4.3 -20.3284944118763\\
4.4 -20.5762288672132\\
4.5 -20.8158219551367\\
4.6 -21.0476699313556\\
4.7 -21.2726000290929\\
4.8 -21.4902422123667\\
4.9 -21.6999553133525\\
5 -21.9010985587644\\
5.1 -22.093040419959\\
5.2 -22.2763610936379\\
5.3 -22.4517398800711\\
5.4 -22.62037846871\\
5.5 -22.7829891214628\\
5.6 -22.9402313094841\\
5.7 -23.0913565278157\\
5.8 -23.2355433817218\\
5.9 -23.3728090687934\\
6 -23.5033904693561\\
6.1 -23.6282058361069\\
6.2 -23.7473813190229\\
6.3 -23.8609590923208\\
6.4 -23.9694983932981\\
6.5 -24.0728480008365\\
6.6 -24.1708237550912\\
6.7 -24.2631062404222\\
6.8 -24.3499622246897\\
6.9 -24.432570725977\\
7 -24.5105178227471\\
7.1 -24.5837120488543\\
7.2 -24.6534137385123\\
7.3 -24.7203022151796\\
7.4 -24.7842378668825\\
7.5 -24.8450559628786\\
7.6 -24.9028468433295\\
7.7 -24.9580271408546\\
7.8 -25.0112082071127\\
7.9 -25.0621386311698\\
8 -25.1103152517339\\
8.1 -25.1559512231117\\
8.2 -25.199888104501\\
8.3 -25.2420878662352\\
8.4 -25.2825161884562\\
8.5 -25.3207459219882\\
8.6 -25.3575685491066\\
8.7 -25.3930946534901\\
8.8 -25.4274555817308\\
8.9 -25.4602281674454\\
9 -25.4914540823763\\
9.1 -25.5206283821832\\
9.2 -25.547736759836\\
9.3 -25.5731571299901\\
9.4 -25.5970169117951\\
9.5 -25.6190275599501\\
9.6 -25.6382378471197\\
9.7 -25.6541227728388\\
9.8 -25.6671564555697\\
9.9 -25.6773323919237\\
10 -25.6848095234396\\
10.1 -25.6899013509316\\
10.2 -25.6936433129098\\
10.3 -25.6966706015003\\
10.4 -25.6982372274163\\
10.5 -25.6985323384519\\
10.6 -25.6987179716878\\
10.7 -25.7001676646817\\
10.8 -25.703121811398\\
10.9 -25.7071586247102\\
11 -25.7134454917271\\
11.1 -25.722606132941\\
11.2 -25.734073042826\\
11.3 -25.7473897841106\\
11.4 -25.7623369160927\\
11.5 -25.7782127902325\\
11.6 -25.7945058658347\\
11.7 -25.8107996796067\\
11.8 -25.8263992249369\\
11.9 -25.8409380345333\\
12 -25.855755933596\\
12.1 -25.871113121097\\
12.2 -25.8864683612753\\
12.3 -25.9013798976084\\
12.4 -25.9152310536055\\
12.5 -25.9280832467691\\
12.6 -25.9400277187045\\
12.7 -25.9520132249029\\
12.8 -25.9643833657592\\
12.9 -25.9776891883402\\
13 -25.9919019039086\\
13.1 -26.0073898340564\\
13.2 -26.0252498749735\\
13.3 -26.0456570250402\\
13.4 -26.0681243299573\\
13.5 -26.0918673689809\\
13.6 -26.1171191724235\\
13.7 -26.1439724341283\\
13.8 -26.1726659471869\\
13.9 -26.2023629125013\\
14 -26.2326466020004\\
14.1 -26.263596902165\\
14.2 -26.2933528246207\\
14.3 -26.3205271148562\\
14.4 -26.3445518311006\\
14.5 -26.3662855357698\\
14.6 -26.3867770346355\\
14.7 -26.407086527581\\
14.8 -26.4276027982428\\
14.9 -26.4484961742021\\
15 -26.4700048724903\\
15.1 -26.4918484913742\\
15.2 -26.5141603940842\\
15.3 -26.5371063310511\\
15.4 -26.5597195472609\\
15.5 -26.5815453421631\\
15.6 -26.6020432514121\\
15.7 -26.6207704762242\\
15.8 -26.6392740586932\\
15.9 -26.6581894282488\\
16 -26.6780973248188\\
16.1 -26.7002302177655\\
16.2 -26.7253223757627\\
16.3 -26.7536405531431\\
16.4 -26.784147597617\\
16.5 -26.8152671904443\\
16.6 -26.8464873162972\\
16.7 -26.8773090065769\\
16.8 -26.9076047357685\\
16.9 -26.9364412820284\\
17 -26.9637598314717\\
17.1 -26.9897788904795\\
17.2 -27.0145863365084\\
17.3 -27.0389213698112\\
17.4 -27.0639418085796\\
17.5 -27.0898648369387\\
17.6 -27.1157391242017\\
17.7 -27.1419551346728\\
17.8 -27.1690643534581\\
17.9 -27.198086133068\\
18 -27.2282688573555\\
18.1 -27.2590623272478\\
18.2 -27.2894379987085\\
18.3 -27.3191757979278\\
18.4 -27.3473155726867\\
18.5 -27.373024610055\\
18.6 -27.3948936344063\\
18.7 -27.4132253009471\\
18.8 -27.4282792688695\\
18.9 -27.4412878343684\\
19 -27.4528529592074\\
19.1 -27.4633986165173\\
19.2 -27.4725104864272\\
19.3 -27.4803851625158\\
19.4 -27.487917881677\\
19.5 -27.4950775412633\\
19.6 -27.5011640823176\\
19.7 -27.5055134593527\\
19.8 -27.5079985825674\\
19.9 -27.5084474728915\\
20 -27.5072848145799\\
};
\addplot [
color=red,
solid,
forget plot
]
table[row sep=crcr]{
0 0\\
0.1 2.16242937353859\\
0.2 3.23616503654397\\
0.3 4.42852400514134\\
0.4 5.0607325363414\\
0.5 6.02207847449357\\
0.6 6.93226132636854\\
0.7 8.09067105492602\\
0.8 8.72849330247151\\
0.9 9.94447498717677\\
1 12.0247244540363\\
1.1 14.1449343002772\\
1.2 14.9956380245352\\
1.3 17.1117732723049\\
1.4 18.8418345248385\\
1.5 22.0533871880126\\
1.6 24.5060483385373\\
1.7 27.1456832458026\\
1.8 30.0220166016806\\
1.9 31.8186486862346\\
2 34.1067083711719\\
2.1 35.5414557479421\\
2.2 36.9621447712173\\
2.3 36.9030500356931\\
2.4 37.4551332846819\\
2.5 35.8581177170258\\
2.6 34.0326517715744\\
2.7 31.1998237928745\\
2.8 27.5737521263412\\
2.9 23.4405336484462\\
3 18.9270814938323\\
3.1 13.9778710454178\\
3.2 7.9768510741144\\
3.3 1.81297102883713\\
3.4 -4.37961618387908\\
3.5 -11.6802125258426\\
3.6 -18.4833735494506\\
3.7 -25.7111342818991\\
3.8 -32.9513591460823\\
3.9 -41.4891109797409\\
4 -49.1751015050914\\
4.1 -57.3032424155261\\
4.2 -66.2870121029962\\
4.3 -75.4627616415459\\
4.4 -84.3106236954472\\
4.5 -94.0269182489377\\
4.6 -104.298622178045\\
4.7 -114.534150184189\\
4.8 -125.342765072053\\
4.9 -134.615637776415\\
5 -146.563517460335\\
5.1 -156.549363387605\\
5.2 -166.966004811048\\
5.3 -176.375769668677\\
5.4 -186.191938802055\\
5.5 -194.387965207023\\
5.6 -202.028834217624\\
5.7 -208.555536237172\\
5.8 -215.787005232566\\
5.9 -221.779469376088\\
6 -225.864898757931\\
6.1 -230.802900468574\\
6.2 -234.575160593545\\
6.3 -238.32105420299\\
6.4 -240.982981191261\\
6.5 -243.1938376653\\
6.6 -246.621081576435\\
6.7 -249.221322753174\\
6.8 -250.668558880734\\
6.9 -253.062979336821\\
7 -255.136025464646\\
7.1 -256.456383680519\\
7.2 -257.881039887776\\
7.3 -258.874735516092\\
7.4 -259.622392126054\\
7.5 -260.255544972388\\
7.6 -260.482735232025\\
7.7 -260.597197580996\\
7.8 -259.772497030651\\
7.9 -259.368485869218\\
8 -258.701088393785\\
8.1 -258.426129528989\\
8.2 -256.844407756586\\
8.3 -257.751031226197\\
8.4 -257.006441354201\\
8.5 -258.173188356767\\
8.6 -259.146274561685\\
8.7 -261.13228790397\\
8.8 -263.293089323632\\
8.9 -266.483043105834\\
9 -270.801330698586\\
9.1 -275.492787739559\\
9.2 -281.594326948896\\
9.3 -286.390422724065\\
9.4 -293.165491855531\\
9.5 -300.325053956324\\
9.6 -308.076662617034\\
9.7 -315.246160260064\\
9.8 -323.76096301743\\
9.9 -331.297277633595\\
10 -339.38052337424\\
10.1 -347.663612027186\\
10.2 -356.217399645239\\
10.3 -365.047445707141\\
10.4 -374.03578571663\\
10.5 -383.08089622967\\
10.6 -392.989917275446\\
10.7 -403.146874363681\\
10.8 -412.893399422435\\
10.9 -423.399267569721\\
11 -434.58986297181\\
11.1 -445.149294003455\\
11.2 -456.471691601519\\
11.3 -467.318882910472\\
11.4 -477.657707096548\\
11.5 -488.075949283633\\
11.6 -497.890249316799\\
11.7 -506.672245682484\\
11.8 -517.05375354867\\
11.9 -523.966489038478\\
12 -531.171778863485\\
12.1 -537.446695687675\\
12.2 -543.217526617176\\
12.3 -548.334532186718\\
12.4 -552.078416548821\\
12.5 -556.219865839699\\
12.6 -559.681802836401\\
12.7 -563.074641113313\\
12.8 -565.00520145793\\
12.9 -567.028561312256\\
13 -570.338753680225\\
13.1 -571.857959860234\\
13.2 -574.244613475831\\
13.3 -576.262930795937\\
13.4 -577.048170102591\\
13.5 -578.586688970058\\
13.6 -579.313929434475\\
13.7 -580.617857540267\\
13.8 -581.146694281242\\
13.9 -580.90939657195\\
14 -580.505054367322\\
14.1 -580.141456384528\\
14.2 -579.830148166241\\
14.3 -578.296020253999\\
14.4 -577.917121595241\\
14.5 -577.018852737743\\
14.6 -576.975802503092\\
14.7 -577.761637934483\\
14.8 -577.30896913977\\
14.9 -577.826340712327\\
15 -580.411941101214\\
15.1 -583.71610040975\\
15.2 -586.929462526915\\
15.3 -590.621076864998\\
15.4 -596.167058910538\\
15.5 -601.15941900852\\
15.6 -607.505809541441\\
15.7 -614.223532695045\\
15.8 -620.913278197709\\
15.9 -628.47256558072\\
16 -635.920495205905\\
16.1 -644.049361825046\\
16.2 -652.007863908491\\
16.3 -660.718504400035\\
16.4 -669.064605067341\\
16.5 -677.982355170945\\
16.6 -686.774348600763\\
16.7 -695.203653745331\\
16.8 -704.846336644345\\
16.9 -714.838076569603\\
17 -724.791475765164\\
17.1 -735.255743956029\\
17.2 -746.180545288898\\
17.3 -756.136757153981\\
17.4 -767.510090785713\\
17.5 -778.517068471549\\
17.6 -790.006691050803\\
17.7 -800.972355866094\\
17.8 -810.341014334094\\
17.9 -821.091502982197\\
18 -830.466373407173\\
18.1 -838.672666333643\\
18.2 -847.031537352942\\
18.3 -854.440998187021\\
18.4 -860.720171533149\\
18.5 -866.584603059147\\
18.6 -871.351696000626\\
18.7 -875.531304201465\\
18.8 -879.813503871734\\
18.9 -883.894065899501\\
19 -886.963720318367\\
19.1 -889.554261757261\\
19.2 -892.84382183615\\
19.3 -894.443612712163\\
19.4 -897.127521666711\\
19.5 -898.539849745931\\
19.6 -901.050836606945\\
19.7 -903.074474893666\\
19.8 -904.384063418918\\
19.9 -905.584701166299\\
20 -906.581093098989\\
};
\addplot [
color=red,
solid,
forget plot
]
table[row sep=crcr]{
0 0\\
0.1 9.15619113077817\\
0.2 11.2844931917514\\
0.3 13.048784936409\\
0.4 10.6628319372733\\
0.5 11.3585295049793\\
0.6 11.3870421656637\\
0.7 11.1124066301764\\
0.8 9.7565009285449\\
0.9 11.2427625700907\\
1 14.6457781552352\\
1.1 15.1941708477576\\
1.2 9.61563221986594\\
1.3 11.9647550249962\\
1.4 10.4175211250457\\
1.5 13.6774443636547\\
1.6 9.67666948401705\\
1.7 7.42305792653568\\
1.8 6.56387252644842\\
1.9 -0.153187327192563\\
2 -2.85514586931076\\
2.1 -9.02789768064707\\
2.2 -11.4857107255933\\
2.3 -19.2771409237584\\
2.4 -20.5713558131171\\
2.5 -29.260202814095\\
2.6 -33.8366634002708\\
2.7 -39.8500286587259\\
2.8 -45.8768265122307\\
2.9 -48.6617182180798\\
3 -52.3458760801269\\
3.1 -55.0703569793234\\
3.2 -60.8891410137463\\
3.3 -64.1533076480318\\
3.4 -66.0404842335137\\
3.5 -70.7941513700595\\
3.6 -71.6145505037797\\
3.7 -74.0827172901878\\
3.8 -77.0539422049967\\
3.9 -82.759007315611\\
4 -82.3824445564542\\
4.1 -84.0633691160999\\
4.2 -88.0860792371148\\
4.3 -90.3758170914893\\
4.4 -89.268250352176\\
4.5 -92.6640105580968\\
4.6 -94.4106820501928\\
4.7 -94.3237544373434\\
4.8 -95.9991137201691\\
4.9 -89.7492867699008\\
5 -96.8870479740256\\
5.1 -91.9317496220144\\
5.2 -91.905407064832\\
5.3 -86.1965781041531\\
5.4 -86.4561470624884\\
5.5 -80.1821975033215\\
5.6 -75.4730013243415\\
5.7 -70.1655329586939\\
5.8 -71.3929111768709\\
5.9 -66.4816692271054\\
6 -58.7656416807241\\
6.1 -57.4418220828476\\
6.2 -52.6815194183206\\
6.3 -51.5595386704128\\
6.4 -46.2658430287922\\
6.5 -42.4843893244604\\
6.6 -45.1582261958167\\
6.7 -41.3650806355481\\
6.8 -36.7628574395749\\
6.9 -37.4990039728233\\
7 -36.0971968623085\\
7.1 -31.3365424807036\\
7.2 -29.8608270658308\\
7.3 -28.2607877483413\\
7.4 -26.826058233943\\
7.5 -26.2485699722202\\
7.6 -24.2032948893464\\
7.7 -23.2800605792343\\
7.8 -20.8656041864008\\
7.9 -22.4727427839508\\
8 -22.2619367327585\\
8.1 -23.5494087542752\\
8.2 -20.8711223146436\\
8.3 -29.1374704656195\\
8.4 -25.6176887743784\\
8.5 -32.4335589450597\\
8.6 -34.0693138637232\\
8.7 -38.9902116477293\\
8.8 -40.8343498523356\\
8.9 -45.0563427400361\\
9 -50.2072128868141\\
9.1 -53.4568095185986\\
9.2 -59.762800322807\\
9.3 -57.3655177833885\\
9.4 -63.5433159955205\\
9.5 -67.953623055041\\
9.6 -72.6196269181175\\
9.7 -71.2443656232329\\
9.8 -76.6447048081849\\
9.9 -75.8996525904013\\
10 -77.5449914597653\\
10.1 -80.0926838477646\\
10.2 -81.8533445144618\\
10.3 -83.8483552833301\\
10.4 -85.1327555407395\\
10.5 -85.7400720124365\\
10.6 -88.4886192869323\\
10.7 -91.3847746142662\\
10.8 -88.8581067957319\\
10.9 -89.7924973489706\\
11 -92.4577988848766\\
11.1 -90.5725685804509\\
11.2 -91.2337659064195\\
11.3 -89.503125240346\\
11.4 -85.9286331825234\\
11.5 -84.3571869246055\\
11.6 -81.3074820861058\\
11.7 -75.9768637479995\\
11.8 -80.2572583395887\\
11.9 -68.44157363103\\
12 -64.6945400174961\\
12.1 -59.9510152467334\\
12.2 -57.3478234486531\\
12.3 -53.8570133083491\\
12.4 -47.3490432222826\\
12.5 -46.5611891351979\\
12.6 -43.3060398782833\\
12.7 -41.4117784404658\\
12.8 -34.7553691899197\\
12.9 -32.2890777779412\\
13 -36.2959042568088\\
13.1 -29.7349771904742\\
13.2 -29.9486882600252\\
13.3 -28.3226462656453\\
13.4 -22.9345125644398\\
13.5 -23.4532300776462\\
13.6 -20.6723331390454\\
13.7 -21.8567998478965\\
13.8 -19.3854152599416\\
13.9 -16.0365832680172\\
14 -14.935711458792\\
14.1 -15.6758122343186\\
14.2 -17.169043393089\\
14.3 -13.5334775845261\\
14.4 -17.9150212782207\\
14.5 -17.7278004954662\\
14.6 -20.7810874794438\\
14.7 -25.9691652382218\\
14.8 -23.3690748692269\\
14.9 -25.7467828441381\\
15 -34.1620621972648\\
15.1 -40.5523081652736\\
15.2 -42.7347080744036\\
15.3 -44.5348788405178\\
15.4 -52.1300101851785\\
15.5 -51.6890434408237\\
15.6 -57.8116833278748\\
15.7 -62.1569611705712\\
15.8 -63.0125494624105\\
15.9 -66.5387660081961\\
16 -68.8097633495106\\
16.1 -72.2518847688413\\
16.2 -73.6970967570849\\
16.3 -77.2717268807127\\
16.4 -77.9607483093105\\
16.5 -80.5939369853\\
16.6 -81.1513049525595\\
16.7 -80.7883398327882\\
16.8 -84.102681515373\\
16.9 -87.2794049420794\\
17 -87.3383343830293\\
17.1 -88.4892943239432\\
17.2 -90.412543641653\\
17.3 -86.1124641158484\\
17.4 -90.3602562552374\\
17.5 -89.1390486856724\\
17.6 -90.2097129038473\\
17.7 -87.1768131677509\\
17.8 -79.6193155394057\\
17.9 -83.5412315338846\\
18 -77.8570595391139\\
18.1 -70.9262153463771\\
18.2 -70.7618032557839\\
18.3 -66.319565835687\\
18.4 -60.2455372695019\\
18.5 -56.6094962785902\\
18.6 -51.79487234767\\
18.7 -48.1207133916129\\
18.8 -46.8239861177688\\
18.9 -45.183953736285\\
19 -40.9959982366479\\
19.1 -36.8108496140782\\
19.2 -38.4402515593604\\
19.3 -31.7455280065735\\
19.4 -32.497395283177\\
19.5 -28.0795253701432\\
19.6 -29.8773834511607\\
19.7 -28.7022611335842\\
19.8 -24.5109167115795\\
19.9 -22.7391715693976\\
20 -21.9484016603322\\
};
\addplot [
color=red,
solid,
forget plot
]
table[row sep=crcr]{
0 0\\
0.1 0.0932093223451834\\
0.2 0.0109574734879643\\
0.3 -0.123612628991892\\
0.4 -0.27447416345894\\
0.5 -0.442071725243783\\
0.6 -0.643008963289661\\
0.7 -0.817880211383888\\
0.8 -1.05349967000463\\
0.9 -1.2907145050128\\
1 -1.53158059475529\\
1.1 -1.79339561789826\\
1.2 -2.0924832787492\\
1.3 -2.36328291021653\\
1.4 -2.67716372667428\\
1.5 -2.94423894134184\\
1.6 -3.22391687290457\\
1.7 -3.48575006837347\\
1.8 -3.76079004693236\\
1.9 -4.03867744981154\\
2 -4.24783227722169\\
2.1 -4.43393523432042\\
2.2 -4.62009344126049\\
2.3 -4.81688587124852\\
2.4 -4.95509094750072\\
2.5 -5.15801651174657\\
2.6 -5.32213658685021\\
2.7 -5.48270198754139\\
2.8 -5.60977329227186\\
2.9 -5.78007222044725\\
3 -5.91979092085827\\
3.1 -6.08059441244632\\
3.2 -6.24904555724557\\
3.3 -6.39981532037138\\
3.4 -6.54201095601449\\
3.5 -6.74100375312779\\
3.6 -6.9072496278279\\
3.7 -7.09822901653888\\
3.8 -7.23513359575905\\
3.9 -7.42501398283934\\
4 -7.55947570638396\\
4.1 -7.71403557141643\\
4.2 -7.88293517734132\\
4.3 -8.03231005414677\\
4.4 -8.17943588215389\\
4.5 -8.31481281317689\\
4.6 -8.47988143617857\\
4.7 -8.63329403825444\\
4.8 -8.79025676648547\\
4.9 -8.91096208704898\\
5 -9.13295986187293\\
5.1 -9.25676206462454\\
5.2 -9.38902725267762\\
5.3 -9.55293635923716\\
5.4 -9.72365190263988\\
5.5 -9.85583576697927\\
5.6 -10.0186341918683\\
5.7 -10.160104224782\\
5.8 -10.3294057949606\\
5.9 -10.4893329831487\\
6 -10.5758115115044\\
6.1 -10.778629266131\\
6.2 -10.9524308939676\\
6.3 -11.1104649841868\\
6.4 -11.2578160366642\\
6.5 -11.4033911616274\\
6.6 -11.5886099748777\\
6.7 -11.7760380153072\\
6.8 -11.8840383154134\\
6.9 -12.0467483517071\\
7 -12.1851874970188\\
7.1 -12.3423514161366\\
7.2 -12.524537759203\\
7.3 -12.6790695013235\\
7.4 -12.8299095818556\\
7.5 -12.974744827769\\
7.6 -13.1369557738661\\
7.7 -13.3213359056906\\
7.8 -13.4595051441696\\
7.9 -13.6078974394647\\
8 -13.7544498161645\\
8.1 -13.9419194857445\\
8.2 -14.04497203396\\
8.3 -14.2418483937816\\
8.4 -14.3813932025298\\
8.5 -14.562141428873\\
8.6 -14.7104606384011\\
8.7 -14.8498912009098\\
8.8 -14.9764256631155\\
8.9 -15.1195922770953\\
9 -15.286462401213\\
9.1 -15.4344567214684\\
9.2 -15.6074476952611\\
9.3 -15.7085355559646\\
9.4 -15.8943567031862\\
9.5 -16.0485724773538\\
9.6 -16.1778297348991\\
9.7 -16.3199648458122\\
9.8 -16.4984423160207\\
9.9 -16.6121270960563\\
10 -16.7708272226345\\
10.1 -16.9004326621666\\
10.2 -17.0425989769409\\
10.3 -17.1800357373312\\
10.4 -17.3138025957626\\
10.5 -17.4391259743253\\
10.6 -17.6009510343142\\
10.7 -17.6959894867359\\
10.8 -17.8054671088332\\
10.9 -17.9568790050238\\
11 -18.1006460798155\\
11.1 -18.1955661791435\\
11.2 -18.3424613828837\\
11.3 -18.4530455578491\\
11.4 -18.568763925361\\
11.5 -18.7026897640808\\
11.6 -18.8197387552145\\
11.7 -18.9285227704326\\
11.8 -19.0915004414852\\
11.9 -19.1515335717115\\
12 -19.298544064103\\
12.1 -19.443803592649\\
12.2 -19.5659232709332\\
12.3 -19.684525805072\\
12.4 -19.7815868313171\\
12.5 -19.9068114825756\\
12.6 -20.0184272594477\\
12.7 -20.1409406695832\\
12.8 -20.228325236975\\
12.9 -20.3398346929219\\
13 -20.4632426943096\\
13.1 -20.5337458390864\\
13.2 -20.6562723240104\\
13.3 -20.7574205458916\\
13.4 -20.8211467467399\\
13.5 -20.9308695529832\\
13.6 -21.0099867314418\\
13.7 -21.1196831757467\\
13.8 -21.2150161560345\\
13.9 -21.3079111548399\\
14 -21.4132032675631\\
14.1 -21.5184127544023\\
14.2 -21.6118059816346\\
14.3 -21.6786897663484\\
14.4 -21.7681554257254\\
14.5 -21.834317079323\\
14.6 -21.9473957500676\\
14.7 -22.0475702628013\\
14.8 -22.0872128931102\\
14.9 -22.1776334010575\\
15 -22.3219921967145\\
15.1 -22.4182412399389\\
15.2 -22.4611470237621\\
15.3 -22.5227092999207\\
15.4 -22.6182830091886\\
15.5 -22.7068056193123\\
15.6 -22.7982683093863\\
15.7 -22.8454164841408\\
15.8 -22.9007614017101\\
15.9 -22.9978509118265\\
16 -23.0476323447376\\
16.1 -23.1259252478569\\
16.2 -23.1802980921113\\
16.3 -23.2659405243247\\
16.4 -23.3215342368017\\
16.5 -23.3983052549326\\
16.6 -23.4714011794528\\
16.7 -23.5056702623895\\
16.8 -23.6100538271361\\
16.9 -23.6739330335111\\
17 -23.7352568866367\\
17.1 -23.8170687315353\\
17.2 -23.8862223699404\\
17.3 -23.9435915744326\\
17.4 -24.0124276884997\\
17.5 -24.0692257660804\\
17.6 -24.1255372239769\\
17.7 -24.191998225\\
17.8 -24.2364413566796\\
17.9 -24.2831571883429\\
18 -24.3312284434767\\
18.1 -24.3986158974051\\
18.2 -24.4418742394751\\
18.3 -24.4901234261639\\
18.4 -24.5455949005947\\
18.5 -24.624591029118\\
18.6 -24.6747313098612\\
18.7 -24.719834068201\\
18.8 -24.7911439887076\\
18.9 -24.8691845841781\\
19 -24.9071821731556\\
19.1 -24.9661306190274\\
19.2 -25.0215980680708\\
19.3 -25.0299531667137\\
19.4 -25.1048925329036\\
19.5 -25.114220352666\\
19.6 -25.1761233464123\\
19.7 -25.1977605402972\\
19.8 -25.2351270151107\\
19.9 -25.2855527655011\\
20 -25.3223271935767\\
};
\end{axis}
\end{tikzpicture}%

%% file: x3.tex
%
%
%
%
\begin{tikzpicture}[scale=0.6]

\begin{axis}[%
width=4.52083333333333in,
height=2in,
scale only axis,
xmin=0,
xmax=20,
xlabel={\Large $t$},
ymin=-10,
ymax=4,
ylabel={\Large $x^{(3)}$}
]
\addplot [
color=black,
solid,
forget plot
]
table[row sep=crcr]{
0 -9.49466532788191\\
0.1 -9.78246689012448\\
0.2 -9.89595409492616\\
0.3 -9.88390442991719\\
0.4 -9.78904110560552\\
0.5 -9.64206229194271\\
0.6 -9.45384698959218\\
0.7 -9.22809797977718\\
0.8 -8.98454229297431\\
0.9 -8.73707581811748\\
1 -8.48213004845782\\
1.1 -8.22377243316994\\
1.2 -7.9674744817158\\
1.3 -7.71771002125771\\
1.4 -7.46907221379485\\
1.5 -7.21876125495095\\
1.6 -6.96512380954095\\
1.7 -6.71106497234056\\
1.8 -6.46450263367544\\
1.9 -6.23080596096036\\
2 -6.01189950454556\\
2.1 -5.80144666230981\\
2.2 -5.59497694655016\\
2.3 -5.39161770822417\\
2.4 -5.18551427649397\\
2.5 -4.9828339078821\\
2.6 -4.79486649039036\\
2.7 -4.62111521607001\\
2.8 -4.45895752509639\\
2.9 -4.30163840307807\\
3 -4.14527460718668\\
3.1 -3.99672923638355\\
3.2 -3.85956047153937\\
3.3 -3.72071328415635\\
3.4 -3.57532810149064\\
3.5 -3.43101617410203\\
3.6 -3.29077335668636\\
3.7 -3.15598046629917\\
3.8 -3.03017772375483\\
3.9 -2.92248139188123\\
4 -2.82053391240741\\
4.1 -2.71501329561216\\
4.2 -2.61396975244166\\
4.3 -2.52034687685026\\
4.4 -2.43591719690431\\
4.5 -2.35592717037749\\
4.6 -2.28341970874036\\
4.7 -2.21400190730484\\
4.8 -2.13774411427004\\
4.9 -2.05566974202381\\
5 -1.96535285163158\\
5.1 -1.87520705331191\\
5.2 -1.79177624684233\\
5.3 -1.71850289759095\\
5.4 -1.654881322476\\
5.5 -1.59933743122221\\
5.6 -1.54364538471089\\
5.7 -1.47710822312676\\
5.8 -1.40712128236096\\
5.9 -1.3380325117966\\
6 -1.27596579304835\\
6.1 -1.22019085329863\\
6.2 -1.16301098428333\\
6.3 -1.11021158320548\\
6.4 -1.05983393859154\\
6.5 -1.00707457425185\\
6.6 -0.95168327582588\\
6.7 -0.893892341932919\\
6.8 -0.846630849398004\\
6.9 -0.803893317722687\\
7 -0.75439774577052\\
7.1 -0.712475669254292\\
7.2 -0.682702946513493\\
7.3 -0.654541603878843\\
7.4 -0.623843226703668\\
7.5 -0.59266375899573\\
7.6 -0.563736559372251\\
7.7 -0.541444463880966\\
7.8 -0.521493665887938\\
7.9 -0.495966987973471\\
8 -0.467753241418612\\
8.1 -0.447234252072017\\
8.2 -0.430720050973807\\
8.3 -0.413881502890704\\
8.4 -0.392788983366494\\
8.5 -0.374367868528579\\
8.6 -0.361537624759796\\
8.7 -0.349837757335892\\
8.8 -0.33589992481202\\
8.9 -0.320395855667589\\
9 -0.302648185540701\\
9.1 -0.280907030961927\\
9.2 -0.262140561441326\\
9.3 -0.246442997548898\\
9.4 -0.230523038526118\\
9.5 -0.20771632757013\\
9.6 -0.175199675514812\\
9.7 -0.144122353960336\\
9.8 -0.116027532839886\\
9.9 -0.0879808197802391\\
10 -0.0616686980105614\\
10.1 -0.0422577138184625\\
10.2 -0.0343277920404141\\
10.3 -0.0238219610493553\\
10.4 -0.00799230541985718\\
10.5 6.29546396786895e-05\\
10.6 -0.00660577426489722\\
10.7 -0.0228652521430201\\
10.8 -0.0342374413592901\\
10.9 -0.0495335234470438\\
11 -0.0774580537181005\\
11.1 -0.104265553386811\\
11.2 -0.124262200543044\\
11.3 -0.142131858089417\\
11.4 -0.155252488743825\\
11.5 -0.161496408656375\\
11.6 -0.163817919233351\\
11.7 -0.160890757767147\\
11.8 -0.149464275679165\\
11.9 -0.14499686120077\\
12 -0.151503059142179\\
12.1 -0.154354333480491\\
12.2 -0.152388396655392\\
12.3 -0.144198271749111\\
12.4 -0.13352648393839\\
12.5 -0.122895627301632\\
12.6 -0.118546542539005\\
12.7 -0.121012730204726\\
12.8 -0.127939625054179\\
12.9 -0.137659293650669\\
13 -0.146916737387314\\
13.1 -0.165507124292604\\
13.2 -0.191685412599701\\
13.3 -0.215815985901783\\
13.4 -0.231404229626209\\
13.5 -0.244632275244338\\
13.6 -0.260025648381595\\
13.7 -0.278291450074343\\
13.8 -0.293466722679304\\
13.9 -0.299481665276852\\
14 -0.307826865619325\\
14.1 -0.306698815267603\\
14.2 -0.286401134245286\\
14.3 -0.255576800940117\\
14.4 -0.226991825407439\\
14.5 -0.209236358734118\\
14.6 -0.202761002480693\\
14.7 -0.203870525572535\\
14.8 -0.206553735449849\\
14.9 -0.212145760788573\\
15 -0.217060843263214\\
15.1 -0.220089065138394\\
15.2 -0.227187042430562\\
15.3 -0.229172827199503\\
15.4 -0.222740042981538\\
15.5 -0.213067756709866\\
15.6 -0.194812319118376\\
15.7 -0.183957164925345\\
15.8 -0.186552485792957\\
15.9 -0.192383564532199\\
16 -0.208446793938234\\
16.1 -0.235169295621464\\
16.2 -0.267621415493333\\
16.3 -0.296869097659685\\
16.4 -0.30979604173862\\
16.5 -0.3125032696467\\
16.6 -0.310486445611463\\
16.7 -0.306649070592339\\
16.8 -0.296692003684214\\
16.9 -0.280342863616755\\
17 -0.266569235872178\\
17.1 -0.253566843456635\\
17.2 -0.243842496488875\\
17.3 -0.245241501429696\\
17.4 -0.25596426005719\\
17.5 -0.259497429740988\\
17.6 -0.259622087179758\\
17.7 -0.264830092091657\\
17.8 -0.280571632212906\\
17.9 -0.297248839867473\\
18 -0.30638184175354\\
18.1 -0.306724234039518\\
18.2 -0.301538372686817\\
18.3 -0.29077196420734\\
18.4 -0.27156187200548\\
18.5 -0.238717953656631\\
18.6 -0.20048604758921\\
18.7 -0.165538177456527\\
18.8 -0.138644247254657\\
18.9 -0.121951042916977\\
19 -0.11068837482149\\
19.1 -0.0988248824654846\\
19.2 -0.0836213491217551\\
19.3 -0.0761249312922807\\
19.4 -0.0742820757461712\\
19.5 -0.0676029339167938\\
19.6 -0.0527096101835024\\
19.7 -0.0345160647800833\\
19.8 -0.0143446532077788\\
19.9 0.00403830674079248\\
20 0.0194872848415263\\
};
\addplot [
color=black,
solid,
forget plot
]
table[row sep=crcr]{
0 -3.94546210528991\\
0.1 -1.90755547969731\\
0.2 -0.433896792668842\\
0.3 0.621958269557203\\
0.4 1.24272698224701\\
0.5 1.67291838789279\\
0.6 2.06849892372499\\
0.7 2.42492333424721\\
0.8 2.44131777642058\\
0.9 2.5009163134967\\
1 2.58933962368082\\
1.1 2.5742652923108\\
1.2 2.54870434896233\\
1.3 2.44748284090233\\
1.4 2.51763151279379\\
1.5 2.48600104917078\\
1.6 2.57793485878947\\
1.7 2.5027672123813\\
1.8 2.42810444874426\\
1.9 2.2507944210102\\
2 2.12973694656289\\
2.1 2.07838178699388\\
2.2 2.04904126434527\\
2.3 2.01638938556784\\
2.4 2.09816475811436\\
2.5 1.95901832279539\\
2.6 1.80489842630377\\
2.7 1.67378318665873\\
2.8 1.57176649556422\\
2.9 1.57198646398295\\
3 1.55360486887011\\
3.1 1.42143107867851\\
3.2 1.32450192793327\\
3.3 1.44415509048653\\
3.4 1.46032133133274\\
3.5 1.42526594698125\\
3.6 1.37953977787421\\
3.7 1.31718840939817\\
3.8 1.20250024905641\\
3.9 0.961661738545566\\
4 1.0701718039748\\
4.1 1.0399914557962\\
4.2 0.982089324775374\\
4.3 0.893246833523819\\
4.4 0.798665057876919\\
4.5 0.799749901190151\\
4.6 0.656353930019114\\
4.7 0.727303336583351\\
4.8 0.793287868827635\\
4.9 0.844286527807503\\
5 0.955017883176982\\
5.1 0.851564307134041\\
5.2 0.817370841709484\\
5.3 0.654986315914535\\
5.4 0.618222298135563\\
5.5 0.497747739603772\\
5.6 0.609581071600365\\
5.7 0.714799871192694\\
5.8 0.685249829005951\\
5.9 0.694893989174125\\
6 0.552517833646131\\
6.1 0.56159908282779\\
6.2 0.580121524823848\\
6.3 0.47998849796415\\
6.4 0.524502763660994\\
6.5 0.529554098327779\\
6.6 0.575076444549843\\
6.7 0.579556586997978\\
6.8 0.375097660016172\\
6.9 0.474002164121588\\
7 0.513131270249755\\
7.1 0.333579614970954\\
7.2 0.264813585152415\\
7.3 0.296368622384991\\
7.4 0.316102421791849\\
7.5 0.307402457838692\\
7.6 0.272407753621916\\
7.7 0.177788564900708\\
7.8 0.218844898606897\\
7.9 0.287819088630632\\
8 0.276549083230185\\
8.1 0.140293995013574\\
8.2 0.187364378672888\\
8.3 0.150945154362395\\
8.4 0.264864813440581\\
8.5 0.110938237983581\\
8.6 0.143811343717834\\
8.7 0.0925512842576581\\
8.8 0.181529642140997\\
8.9 0.130825916992759\\
9 0.219408132680212\\
9.1 0.215260241643213\\
9.2 0.162396880079017\\
9.3 0.151821283887101\\
9.4 0.165623604877047\\
9.5 0.284208661604742\\
9.6 0.361712479448933\\
9.7 0.264187266747488\\
9.8 0.295669265397522\\
9.9 0.266266594046108\\
10 0.259857948775239\\
10.1 0.134308795300449\\
10.2 0.0293971245910625\\
10.3 0.173355115151194\\
10.4 0.144419611126375\\
10.5 0.0226337167689949\\
10.6 -0.147405251554555\\
10.7 -0.176081531474203\\
10.8 -0.0571143970348403\\
10.9 -0.23944665503771\\
11 -0.314815126646967\\
11.1 -0.225356396498127\\
11.2 -0.176674950044665\\
11.3 -0.180242595213109\\
11.4 -0.0866269615605944\\
11.5 -0.0404534943525236\\
11.6 -0.00757984485495627\\
11.7 0.0625709011253067\\
11.8 0.160859830409818\\
11.9 -0.0605293334808777\\
12 -0.0690602824338126\\
12.1 0.00822260968702492\\
12.2 0.0299769544833168\\
12.3 0.12875602232741\\
12.4 0.0866070990554138\\
12.5 0.123967256371281\\
12.6 -0.0293985886395464\\
12.7 -0.0203366837688662\\
12.8 -0.113436356846188\\
12.9 -0.082347851038083\\
13 -0.101681387452234\\
13.1 -0.26181935094325\\
13.2 -0.261334886195322\\
13.3 -0.222799465810263\\
13.4 -0.0950842723439596\\
13.5 -0.165728375014899\\
13.6 -0.143017113457429\\
13.7 -0.218238220863764\\
13.8 -0.0913515888125077\\
13.9 -0.0318202476723529\\
14 -0.130039609515821\\
14.1 0.139138606008854\\
14.2 0.260420155695948\\
14.3 0.351028323694072\\
14.4 0.226418632497351\\
14.5 0.1330577762601\\
14.6 0.00284464324304299\\
14.7 -0.0236913817674724\\
14.8 -0.0296314953954284\\
14.9 -0.0796194484439828\\
15 -0.0215028380071229\\
15.1 -0.0381783898415022\\
15.2 -0.100548157716344\\
15.3 0.0531876761545332\\
15.4 0.0743062395607881\\
15.5 0.116853605261658\\
15.6 0.241715472451845\\
15.7 -0.0121162627947141\\
15.8 -0.0384326690687668\\
15.9 -0.0762054130206893\\
16 -0.236772793043768\\
16.1 -0.294356620571307\\
16.2 -0.351301690973397\\
16.3 -0.23878444931603\\
16.4 -0.0299679348108449\\
16.5 -0.0244091771074442\\
16.6 0.0604729269257554\\
16.7 0.0183160891183682\\
16.8 0.172937455304442\\
16.9 0.15468479320706\\
17 0.122181765306893\\
17.1 0.136913915376543\\
17.2 0.0611928126148944\\
17.3 -0.0819984039588565\\
17.4 -0.129886669462046\\
17.5 0.0502840406945772\\
17.6 -0.0478729799258598\\
17.7 -0.0558043282561982\\
17.8 -0.249110417556958\\
17.9 -0.0920024032253548\\
18 -0.0905768167220211\\
18.1 0.0754433311491031\\
18.2 0.0304353574792883\\
18.3 0.177375177205375\\
18.4 0.205121237645062\\
18.5 0.439504924642218\\
18.6 0.329967346216954\\
18.7 0.36657987313761\\
18.8 0.180161173834821\\
18.9 0.154696798989588\\
19 0.0743802516135148\\
19.1 0.158491471263763\\
19.2 0.145952365720176\\
19.3 0.0106104316674466\\
19.4 0.0254737076745031\\
19.5 0.10407259128538\\
19.6 0.18929008012101\\
19.7 0.174992073479409\\
19.8 0.225574252126195\\
19.9 0.145764797094303\\
20 0.16213982684233\\
};
\addplot [
color=red,
solid,
forget plot
]
table[row sep=crcr]{
0 0\\
0.1 -0.164293581422115\\
0.2 -0.329955080092184\\
0.3 -0.498728398305213\\
0.4 -0.658408736028834\\
0.5 -0.807472237269923\\
0.6 -0.938405551576494\\
0.7 -1.06612235763702\\
0.8 -1.18703593252343\\
0.9 -1.29388312427739\\
1 -1.39129994274358\\
1.1 -1.47853508947096\\
1.2 -1.55314101139906\\
1.3 -1.61662096569542\\
1.4 -1.67775568609183\\
1.5 -1.73367552337999\\
1.6 -1.77587911579793\\
1.7 -1.82180522682931\\
1.8 -1.85497846119192\\
1.9 -1.88668543083674\\
2 -1.89582102061524\\
2.1 -1.91243746975094\\
2.2 -1.92210473047168\\
2.3 -1.92127481655896\\
2.4 -1.91805217749514\\
2.5 -1.90653897099706\\
2.6 -1.89652964904106\\
2.7 -1.88375700666871\\
2.8 -1.87337122350036\\
2.9 -1.86196458095918\\
3 -1.83876349666601\\
3.1 -1.811246169933\\
3.2 -1.79681087053049\\
3.3 -1.77387040904329\\
3.4 -1.73769389432358\\
3.5 -1.71704644709509\\
3.6 -1.68138578361206\\
3.7 -1.64475408206702\\
3.8 -1.60488277747493\\
3.9 -1.58921638026847\\
4 -1.55325428643253\\
4.1 -1.52242240399945\\
4.2 -1.48457176553638\\
4.3 -1.46260600605896\\
4.4 -1.42356783851892\\
4.5 -1.39285703561555\\
4.6 -1.36881310747985\\
4.7 -1.33908555394088\\
4.8 -1.30167492608706\\
4.9 -1.26534461502353\\
5 -1.23126309034554\\
5.1 -1.19803941506477\\
5.2 -1.15923976792311\\
5.3 -1.12890354150006\\
5.4 -1.10096669235716\\
5.5 -1.05532103871937\\
5.6 -1.0243597331127\\
5.7 -1.00071268892577\\
5.8 -0.976209707835312\\
5.9 -0.957568677439602\\
6 -0.943988414348219\\
6.1 -0.90651752591408\\
6.2 -0.87031847405507\\
6.3 -0.852304393074329\\
6.4 -0.83204895735219\\
6.5 -0.8034340243185\\
6.6 -0.778361699283991\\
6.7 -0.75216689854196\\
6.8 -0.73175936171765\\
6.9 -0.707047158403587\\
7 -0.673818222308744\\
7.1 -0.652133683910872\\
7.2 -0.623261916114641\\
7.3 -0.606015661484927\\
7.4 -0.578332344742907\\
7.5 -0.563216869143288\\
7.6 -0.538018496800642\\
7.7 -0.527546650348103\\
7.8 -0.512755522533837\\
7.9 -0.487864046713286\\
8 -0.465048354868528\\
8.1 -0.449918176874415\\
8.2 -0.436919832246444\\
8.3 -0.424741014789467\\
8.4 -0.412800741121586\\
8.5 -0.395691774051307\\
8.6 -0.389023953982503\\
8.7 -0.366598237792157\\
8.8 -0.351422795853861\\
8.9 -0.336484694614274\\
9 -0.328252945283328\\
9.1 -0.308964414834636\\
9.2 -0.299413224545741\\
9.3 -0.285437235418889\\
9.4 -0.282502287678245\\
9.5 -0.267819124385408\\
9.6 -0.267764969373555\\
9.7 -0.250092304364789\\
9.8 -0.231176629812804\\
9.9 -0.228315694998984\\
10 -0.220770429000528\\
10.1 -0.217391404961465\\
10.2 -0.208372239960337\\
10.3 -0.20110708945055\\
10.4 -0.185057343120724\\
10.5 -0.179069038530487\\
10.6 -0.179373998413458\\
10.7 -0.160999661919025\\
10.8 -0.142032508046354\\
10.9 -0.13361812333557\\
11 -0.123480038119644\\
11.1 -0.111951683580494\\
11.2 -0.116390424790423\\
11.3 -0.122337576583773\\
11.4 -0.118095577617589\\
11.5 -0.10759208998385\\
11.6 -0.107805186484353\\
11.7 -0.107160186736703\\
11.8 -0.11353326475526\\
11.9 -0.116620419393708\\
12 -0.110395358852333\\
12.1 -0.112511690037413\\
12.2 -0.108981872415938\\
12.3 -0.104533936434015\\
12.4 -0.104352996444568\\
12.5 -0.0979341374139975\\
12.6 -0.0949003543908858\\
12.7 -0.0885560256556069\\
12.8 -0.0861827231363143\\
12.9 -0.0864253820123208\\
13 -0.0771714690889432\\
13.1 -0.0637982653402391\\
13.2 -0.0654058928532819\\
13.3 -0.0590087084870132\\
13.4 -0.0531183207475076\\
13.5 -0.0455988559665347\\
13.6 -0.0471814904604508\\
13.7 -0.0608400106390601\\
13.8 -0.0626545886142587\\
13.9 -0.06431501989647\\
14 -0.0717161075214775\\
14.1 -0.0729683777462851\\
14.2 -0.0685309434381861\\
14.3 -0.0731172076509854\\
14.4 -0.0699437361441875\\
14.5 -0.0673239729797199\\
14.6 -0.0717582028438631\\
14.7 -0.0752101466466868\\
14.8 -0.0701478713308425\\
14.9 -0.0727969593393085\\
15 -0.0734645665932946\\
15.1 -0.0762712379724523\\
15.2 -0.0658692469861096\\
15.3 -0.0632502294259102\\
15.4 -0.0661655686023115\\
15.5 -0.0628827530868422\\
15.6 -0.0633145239220446\\
15.7 -0.06501330031587\\
15.8 -0.0639153937950179\\
15.9 -0.0629913962078283\\
16 -0.0567309012940927\\
16.1 -0.0585195071472103\\
16.2 -0.0534944735488467\\
16.3 -0.0584793859342854\\
16.4 -0.0688263513060006\\
16.5 -0.0687152516824661\\
16.6 -0.0683140829244264\\
16.7 -0.0765278995739417\\
16.8 -0.0705227637073836\\
16.9 -0.0700809331128156\\
17 -0.0692989199605028\\
17.1 -0.0715369472803962\\
17.2 -0.0710869028535178\\
17.3 -0.0787335855960045\\
17.4 -0.0830853249466721\\
17.5 -0.081528405276489\\
17.6 -0.075539623333132\\
17.7 -0.0782505630763044\\
17.8 -0.0879765325879198\\
17.9 -0.0915066221108366\\
18 -0.0969612929732591\\
18.1 -0.0978178563077843\\
18.2 -0.104649250110537\\
18.3 -0.105032680437019\\
18.4 -0.0978367298329734\\
18.5 -0.0903423663517661\\
18.6 -0.0893036036176632\\
18.7 -0.078459907811826\\
18.8 -0.0722505367212807\\
18.9 -0.073283204388531\\
19 -0.0727301932894913\\
19.1 -0.0715275194027888\\
19.2 -0.0710650918986245\\
19.3 -0.070922080906571\\
19.4 -0.0674192668237199\\
19.5 -0.0687905333994105\\
19.6 -0.0613370815953324\\
19.7 -0.0614492833620008\\
19.8 -0.0561009253841366\\
19.9 -0.0552345811772827\\
20 -0.0486555948583379\\
};
\addplot [
color=red,
solid,
forget plot
]
table[row sep=crcr]{
0 0\\
0.1 0.00973705334551013\\
0.2 0.0312775983841003\\
0.3 0.0617905225532756\\
0.4 0.0984975409682912\\
0.5 0.138918476420122\\
0.6 0.180727766800168\\
0.7 0.222884986337951\\
0.8 0.264806226811727\\
0.9 0.30553013230878\\
1 0.34442498224771\\
1.1 0.381101859436337\\
1.2 0.415106212364365\\
1.3 0.44612417268828\\
1.4 0.47448707548328\\
1.5 0.500507073486891\\
1.6 0.523832403460366\\
1.7 0.54500886773037\\
1.8 0.563932391637525\\
1.9 0.580751883252964\\
2 0.594628124742442\\
2.1 0.606156371383982\\
2.2 0.615732274774548\\
2.3 0.623047690027912\\
2.4 0.628260766921635\\
2.5 0.631355633539886\\
2.6 0.632743050233806\\
2.7 0.632740185865693\\
2.8 0.631773532371585\\
2.9 0.630096561846892\\
3 0.627163980486278\\
3.1 0.622723440723494\\
3.2 0.617825228069116\\
3.3 0.612436970391577\\
3.4 0.605668483719139\\
3.5 0.598440930088586\\
3.6 0.590399379723543\\
3.7 0.581290098115853\\
3.8 0.571169347839634\\
3.9 0.56164922557545\\
4 0.552023558121977\\
4.1 0.54207703206997\\
4.2 0.531612752802757\\
4.3 0.521519065732591\\
4.4 0.511114143237442\\
4.5 0.500512608887859\\
4.6 0.490386528366935\\
4.7 0.480455951153098\\
4.8 0.470060927493051\\
4.9 0.459177743084284\\
5 0.448095698305743\\
5.1 0.436976895059248\\
5.2 0.425521594716451\\
5.3 0.414168850085375\\
5.4 0.403263840507335\\
5.5 0.391702709328116\\
5.6 0.380050473243236\\
5.7 0.369164042969671\\
5.8 0.358991655801675\\
5.9 0.349673983151443\\
6 0.341449111000525\\
6.1 0.332750882369313\\
6.2 0.323132083143525\\
6.3 0.313945669307541\\
6.4 0.305438687547406\\
6.5 0.296878475117201\\
6.6 0.288263490760835\\
6.7 0.279636861924475\\
6.8 0.271313859739267\\
6.9 0.263117346000751\\
7 0.254393036633216\\
7.1 0.245731777391002\\
7.2 0.237001432832753\\
7.3 0.228718271896994\\
7.4 0.220460899796014\\
7.5 0.212687721188158\\
7.6 0.205013502231712\\
7.7 0.198015444138648\\
7.8 0.191660828711144\\
7.9 0.185066944733925\\
8 0.178145552001233\\
8.1 0.171493180978739\\
8.2 0.165375485716393\\
8.3 0.159761319589374\\
8.4 0.154557550093948\\
8.5 0.149356343527764\\
8.6 0.144638013506701\\
8.7 0.139623462967465\\
8.8 0.134394763582975\\
8.9 0.129217244108105\\
9 0.12448311901146\\
9.1 0.11960178957103\\
9.2 0.114890611483814\\
9.3 0.110305766040139\\
9.4 0.106352112926359\\
9.5 0.102467295462373\\
9.6 0.0991863322083896\\
9.7 0.095699564873989\\
9.8 0.0915096426349861\\
9.9 0.0877197495972547\\
10 0.084381907711878\\
10.1 0.0814999063582581\\
10.2 0.0787326998988854\\
10.3 0.0760042278209741\\
10.4 0.0728296977642865\\
10.5 0.0696835868052858\\
10.6 0.0672040889204344\\
10.7 0.0642788930041964\\
10.8 0.0604844987762407\\
10.9 0.0566558080175128\\
11 0.0529834751993564\\
11.1 0.0492995281070427\\
11.2 0.0465197626502757\\
11.3 0.0449316395948696\\
11.4 0.0436626018223507\\
11.5 0.0419825750325969\\
11.6 0.0404654134341676\\
11.7 0.0393079730590254\\
11.8 0.0388120152073266\\
11.9 0.0387991488414817\\
12 0.0385014951019972\\
12.1 0.0382386995135589\\
12.2 0.0378916046693859\\
12.3 0.0372748841212365\\
12.4 0.0366866321479336\\
12.5 0.0358635660446735\\
12.6 0.0348981784033735\\
12.7 0.0337218113741439\\
12.8 0.0325358670296294\\
12.9 0.0316060223927832\\
13 0.0303724214858094\\
13.1 0.0284104721038069\\
13.2 0.0266881292791789\\
13.3 0.0250736547694039\\
13.4 0.0233537019766877\\
13.5 0.0214715130860294\\
13.6 0.0199657926621724\\
13.7 0.0197036256962212\\
13.8 0.0199867962410586\\
13.9 0.0203458057806067\\
14 0.0210969378036027\\
14.1 0.0219299713862697\\
14.2 0.0223202048574304\\
14.3 0.0227637857084062\\
14.4 0.0230278022292873\\
14.5 0.0229807024104028\\
14.6 0.0231372819155097\\
14.7 0.0235777405405094\\
14.8 0.0236965354341867\\
14.9 0.0238049843429679\\
15 0.0239968717365255\\
15.1 0.0243244346932629\\
15.2 0.024024112466009\\
15.3 0.0233649334031981\\
15.4 0.0229795325593349\\
15.5 0.0225775970538067\\
15.6 0.0222174058358112\\
15.7 0.0220619640000369\\
15.8 0.0219256603613439\\
15.9 0.0217395006927924\\
16 0.0212027727472867\\
16.1 0.0207404171499509\\
16.2 0.0201461973903862\\
16.3 0.0198664236903011\\
16.4 0.0204111160072628\\
16.5 0.0210895871309699\\
16.6 0.0215666266578014\\
16.7 0.0223984270983225\\
16.8 0.0228811784309553\\
16.9 0.0230514078911793\\
17 0.0231189117722953\\
17.1 0.0232808716903628\\
17.2 0.0234343526648541\\
17.3 0.0239894896165462\\
17.4 0.0248657813694607\\
17.5 0.0255418226012114\\
17.6 0.025645409475489\\
17.7 0.0257209531158766\\
17.8 0.0264274776339469\\
17.9 0.0274244876366427\\
18 0.0285849070270275\\
18.1 0.0296452890926439\\
18.2 0.030862061788883\\
18.3 0.031972942507989\\
18.4 0.0323804118816543\\
18.5 0.0320440499389821\\
18.6 0.0315307124748238\\
18.7 0.0304764482942906\\
18.8 0.0290307992354802\\
18.9 0.0278506551365946\\
19 0.0269695368251029\\
19.1 0.0262294971423596\\
19.2 0.0256208811043031\\
19.3 0.0251474155247224\\
19.4 0.0245832931185505\\
19.5 0.0241522555757778\\
19.6 0.023427254836607\\
19.7 0.0226949503772796\\
19.8 0.0218368661461515\\
19.9 0.0210039242089158\\
20 0.0199709276401035\\
};
\end{axis}
\end{tikzpicture}%

%% file: x4.tex
%
%
%
%
\begin{tikzpicture}[scale=0.6]

\begin{axis}[%
width=4.52083333333333in,
height=2in,
scale only axis,
xmin=0,
xmax=20,
xlabel={\Large $t$},
ymin=-30,
ymax=30,
ylabel={\Large $x^{(4)}$}
]
\addplot [
color=black,
solid,
forget plot
]
table[row sep=crcr]{
0 5.60326300095945\\
0.1 4.63771027544997\\
0.2 3.65256276478974\\
0.3 2.66269113557899\\
0.4 1.67852727960912\\
0.5 0.706614155234748\\
0.6 -0.248510442190662\\
0.7 -1.18290422172118\\
0.8 -2.09354977789726\\
0.9 -2.9796801798533\\
1 -3.84071395513887\\
1.1 -4.67599642653087\\
1.2 -5.48553739323113\\
1.3 -6.26971227551187\\
1.4 -7.02910966325068\\
1.5 -7.76347490893413\\
1.6 -8.47274556618321\\
1.7 -9.1564923432752\\
1.8 -9.81520848486652\\
1.9 -10.4498262557182\\
2 -11.0618606909934\\
2.1 -11.6524851727955\\
2.2 -12.222281849465\\
2.3 -12.7715843239338\\
2.4 -13.3005088928697\\
2.5 -13.8088104167697\\
2.6 -14.2975670962079\\
2.7 -14.7682569919963\\
2.8 -15.222175660858\\
2.9 -15.6602055748912\\
3 -16.0825358627565\\
3.1 -16.4895259949365\\
3.2 -16.8822597565041\\
3.3 -17.2613729647734\\
3.4 -17.6261884275149\\
3.5 -17.9764764073314\\
3.6 -18.3125277711169\\
3.7 -18.6348135158881\\
3.8 -18.944025926814\\
3.9 -19.2414584061966\\
4 -19.5286994336934\\
4.1 -19.8054516333897\\
4.2 -20.0718525562149\\
4.3 -20.3284944118763\\
4.4 -20.5762288672132\\
4.5 -20.8158219551367\\
4.6 -21.0476699313556\\
4.7 -21.2726000290929\\
4.8 -21.4902422123667\\
4.9 -21.6999553133525\\
5 -21.9010985587644\\
5.1 -22.093040419959\\
5.2 -22.2763610936379\\
5.3 -22.4517398800711\\
5.4 -22.62037846871\\
5.5 -22.7829891214628\\
5.6 -22.9402313094841\\
5.7 -23.0913565278157\\
5.8 -23.2355433817218\\
5.9 -23.3728090687934\\
6 -23.5033904693561\\
6.1 -23.6282058361069\\
6.2 -23.7473813190229\\
6.3 -23.8609590923208\\
6.4 -23.9694983932981\\
6.5 -24.0728480008365\\
6.6 -24.1708237550912\\
6.7 -24.2631062404222\\
6.8 -24.3499622246897\\
6.9 -24.432570725977\\
7 -24.5105178227471\\
7.1 -24.5837120488543\\
7.2 -24.6534137385123\\
7.3 -24.7203022151796\\
7.4 -24.7842378668825\\
7.5 -24.8450559628786\\
7.6 -24.9028468433295\\
7.7 -24.9580271408546\\
7.8 -25.0112082071127\\
7.9 -25.0621386311698\\
8 -25.1103152517339\\
8.1 -25.1559512231117\\
8.2 -25.199888104501\\
8.3 -25.2420878662352\\
8.4 -25.2825161884562\\
8.5 -25.3207459219882\\
8.6 -25.3575685491066\\
8.7 -25.3930946534901\\
8.8 -25.4274555817308\\
8.9 -25.4602281674454\\
9 -25.4914540823763\\
9.1 -25.5206283821832\\
9.2 -25.547736759836\\
9.3 -25.5731571299901\\
9.4 -25.5970169117951\\
9.5 -25.6190275599501\\
9.6 -25.6382378471197\\
9.7 -25.6541227728388\\
9.8 -25.6671564555697\\
9.9 -25.6773323919237\\
10 -25.6848095234396\\
10.1 -25.6899013509316\\
10.2 -25.6936433129098\\
10.3 -25.6966706015003\\
10.4 -25.6982372274163\\
10.5 -25.6985323384519\\
10.6 -25.6987179716878\\
10.7 -25.7001676646817\\
10.8 -25.703121811398\\
10.9 -25.7071586247102\\
11 -25.7134454917271\\
11.1 -25.722606132941\\
11.2 -25.734073042826\\
11.3 -25.7473897841106\\
11.4 -25.7623369160927\\
11.5 -25.7782127902325\\
11.6 -25.7945058658347\\
11.7 -25.8107996796067\\
11.8 -25.8263992249369\\
11.9 -25.8409380345333\\
12 -25.855755933596\\
12.1 -25.871113121097\\
12.2 -25.8864683612753\\
12.3 -25.9013798976084\\
12.4 -25.9152310536055\\
12.5 -25.9280832467691\\
12.6 -25.9400277187045\\
12.7 -25.9520132249029\\
12.8 -25.9643833657592\\
12.9 -25.9776891883402\\
13 -25.9919019039086\\
13.1 -26.0073898340564\\
13.2 -26.0252498749735\\
13.3 -26.0456570250402\\
13.4 -26.0681243299573\\
13.5 -26.0918673689809\\
13.6 -26.1171191724235\\
13.7 -26.1439724341283\\
13.8 -26.1726659471869\\
13.9 -26.2023629125013\\
14 -26.2326466020004\\
14.1 -26.263596902165\\
14.2 -26.2933528246207\\
14.3 -26.3205271148562\\
14.4 -26.3445518311006\\
14.5 -26.3662855357698\\
14.6 -26.3867770346355\\
14.7 -26.407086527581\\
14.8 -26.4276027982428\\
14.9 -26.4484961742021\\
15 -26.4700048724903\\
15.1 -26.4918484913742\\
15.2 -26.5141603940842\\
15.3 -26.5371063310511\\
15.4 -26.5597195472609\\
15.5 -26.5815453421631\\
15.6 -26.6020432514121\\
15.7 -26.6207704762242\\
15.8 -26.6392740586932\\
15.9 -26.6581894282488\\
16 -26.6780973248188\\
16.1 -26.7002302177655\\
16.2 -26.7253223757627\\
16.3 -26.7536405531431\\
16.4 -26.784147597617\\
16.5 -26.8152671904443\\
16.6 -26.8464873162972\\
16.7 -26.8773090065769\\
16.8 -26.9076047357685\\
16.9 -26.9364412820284\\
17 -26.9637598314717\\
17.1 -26.9897788904795\\
17.2 -27.0145863365084\\
17.3 -27.0389213698112\\
17.4 -27.0639418085796\\
17.5 -27.0898648369387\\
17.6 -27.1157391242017\\
17.7 -27.1419551346728\\
17.8 -27.1690643534581\\
17.9 -27.198086133068\\
18 -27.2282688573555\\
18.1 -27.2590623272478\\
18.2 -27.2894379987085\\
18.3 -27.3191757979278\\
18.4 -27.3473155726867\\
18.5 -27.373024610055\\
18.6 -27.3948936344063\\
18.7 -27.4132253009471\\
18.8 -27.4282792688695\\
18.9 -27.4412878343684\\
19 -27.4528529592074\\
19.1 -27.4633986165173\\
19.2 -27.4725104864272\\
19.3 -27.4803851625158\\
19.4 -27.487917881677\\
19.5 -27.4950775412633\\
19.6 -27.5011640823176\\
19.7 -27.5055134593527\\
19.8 -27.5079985825674\\
19.9 -27.5084474728915\\
20 -27.5072848145799\\
};
\addplot [
color=black,
solid,
forget plot
]
table[row sep=crcr]{
0 3.76855682189071\\
0.1 4.63288277278549\\
0.2 5.0474079990485\\
0.3 5.13472157942113\\
0.4 5.03710475867448\\
0.5 4.90238039626409\\
0.6 4.87238458101159\\
0.7 5.07133274909823\\
0.8 5.59246549695875\\
0.9 6.48570955881835\\
1 7.7562388284103\\
1.1 9.36362078453649\\
1.2 11.2231820523185\\
1.3 13.2135683519054\\
1.4 15.1874837575319\\
1.5 16.9866744274854\\
1.6 18.453830910407\\
1.7 19.4476872550379\\
1.8 19.8531687636954\\
1.9 19.5929259620623\\
2 18.6344722955143\\
2.1 16.9956123161087\\
2.2 14.7441393795585\\
2.3 11.9919474727879\\
2.4 8.8859010319232\\
2.5 5.59599110161316\\
2.6 2.29788889585666\\
2.7 -0.839645938640634\\
2.8 -3.66728373722254\\
2.9 -6.06657015228515\\
3 -7.95853049505184\\
3.1 -9.31105553951392\\
3.2 -10.1414566599414\\
3.3 -10.5111198164129\\
3.4 -10.5169051639436\\
3.5 -10.2847269418511\\
3.6 -9.9572160695565\\
3.7 -9.67906075488373\\
3.8 -9.5829629148187\\
3.9 -9.77738837938513\\
4 -10.3351372127495\\
4.1 -11.2815000092292\\
4.2 -12.5958358607931\\
4.3 -14.2128145727146\\
4.4 -16.0272855164279\\
4.5 -17.9025109533051\\
4.6 -19.6814970924824\\
4.7 -21.2024393534309\\
4.8 -22.3099263004999\\
4.9 -22.8708705128384\\
5 -22.7873399468227\\
5.1 -22.0058721254105\\
5.2 -20.5252810224861\\
5.3 -18.3952344145384\\
5.4 -15.7133956541628\\
5.5 -12.6155225421918\\
5.6 -9.26515300770659\\
5.7 -5.83740355806309\\
5.8 -2.50726762715088\\
5.9 0.562826892995165\\
6 3.23739056411654\\
6.1 5.41771482636173\\
6.2 7.05227425889251\\
6.3 8.13851925650815\\
6.4 8.7204939423949\\
6.5 8.88595826442202\\
6.6 8.7558108720831\\
6.7 8.47211077981456\\
6.8 8.18262891576635\\
6.9 8.02582042041685\\
7 8.12237085616025\\
7.1 8.55954316462366\\
7.2 9.38038766382741\\
7.3 10.583004999015\\
7.4 12.1205431257726\\
7.5 13.9031987805137\\
7.6 15.8051933938392\\
7.7 17.6755415675706\\
7.8 19.3516548466346\\
7.9 20.6759608579131\\
8 21.5088303047018\\
8.1 21.7393019655816\\
8.2 21.2955596823833\\
8.3 20.1549057234359\\
8.4 18.3446805431512\\
8.5 15.9413260452236\\
8.6 13.0612093881003\\
8.7 9.85501446948694\\
8.8 6.4939919706927\\
8.9 3.15694526830653\\
9 0.0134340780178818\\
9.1 -2.78810343216154\\
9.2 -5.1343735201056\\
9.3 -6.95578924098915\\
9.4 -8.22998334444815\\
9.5 -8.98206121753039\\
9.6 -9.28051698064025\\
9.7 -9.23200070728769\\
9.8 -8.97163594493369\\
9.9 -8.64668854733461\\
10 -8.40343691414689\\
10.1 -8.37290482062973\\
10.2 -8.65914650612616\\
10.3 -9.32662326751877\\
10.4 -10.3915579142409\\
10.5 -11.8243577416577\\
10.6 -13.5516675720798\\
10.7 -15.4607543774671\\
10.8 -17.4063755575264\\
10.9 -19.2244115863817\\
11 -20.7508961130938\\
11.1 -21.832369598318\\
11.2 -22.3384163832454\\
11.3 -22.1762730537997\\
11.4 -21.2998842381744\\
11.5 -19.7132303642871\\
11.6 -17.4721042421981\\
11.7 -14.6798515218972\\
11.8 -11.4784114596877\\
11.9 -8.03798468635277\\
12 -4.54572348486396\\
12.1 -1.1846047884678\\
12.2 1.88111253468067\\
12.3 4.51737364083103\\
12.4 6.63139339952422\\
12.5 8.17678593738466\\
12.6 9.15755587529508\\
12.7 9.62471410151172\\
12.8 9.67329852196664\\
12.9 9.43096545995383\\
13 9.04686252395899\\
13.1 8.67510717559002\\
13.2 8.4592100477325\\
13.3 8.52150854304258\\
13.4 8.95072281866293\\
13.5 9.79231184304909\\
13.6 11.0409158593532\\
13.7 12.6424298861443\\
13.8 14.4977225353305\\
13.9 16.4734712394304\\
14 18.4107228739399\\
14.1 20.1387325777568\\
14.2 21.4949873665987\\
14.3 22.3348782388507\\
14.4 22.5440166694629\\
14.5 22.0476025346309\\
14.6 20.8201811003831\\
14.7 18.8890800648219\\
14.8 16.334237501187\\
14.9 13.2811338685312\\
15 9.89006448473291\\
15.1 6.34411699815939\\
15.2 2.83215361324538\\
15.3 -0.466014507224786\\
15.4 -3.3920483207699\\
15.5 -5.82518889237865\\
15.6 -7.69051341253226\\
15.7 -8.96462816065677\\
15.8 -9.68008156751677\\
15.9 -9.91514184217968\\
16 -9.78699873545713\\
16.1 -9.44124850469284\\
16.2 -9.03516542438547\\
16.3 -8.72236800414421\\
16.4 -8.63625431187762\\
16.5 -8.87864554977913\\
16.6 -9.5130598788687\\
16.7 -10.5561796652557\\
16.8 -11.9763688248325\\
16.9 -13.6935919172239\\
17 -15.5894863843417\\
17.1 -17.5164221176728\\
17.2 -19.3096462188262\\
17.3 -20.8034559787973\\
17.4 -21.845522995389\\
17.5 -22.3078922417908\\
17.6 -22.0982991967215\\
17.7 -21.1739479783284\\
17.8 -19.5436029943996\\
17.9 -17.2680706349417\\
18 -14.4511230207729\\
18.1 -11.2351618409473\\
18.2 -7.7883265001121\\
18.3 -4.29328417331342\\
18.4 -0.929151908209357\\
18.5 2.14138730929902\\
18.6 4.78652212230668\\
18.7 6.91103569953322\\
18.8 8.46686650268407\\
18.9 9.45403192702728\\
19 9.92250386961214\\
19.1 9.96554781884435\\
19.2 9.71137274726781\\
19.3 9.30746743458671\\
19.4 8.90602654787609\\
19.5 8.65248268979662\\
19.6 8.67091531481662\\
19.7 9.05049049976931\\
19.8 9.83553880246664\\
19.9 11.0219397320987\\
20 12.5555591096769\\
};
\addplot [
color=black,
solid,
forget plot
]
table[row sep=crcr]{
0 8.4068230545413\\
0.1 7.4090267688342\\
0.2 6.46252120547105\\
0.3 5.65206223673188\\
0.4 5.02975173649963\\
0.5 4.61309724070473\\
0.6 4.38567312551019\\
0.7 4.29992138013829\\
0.8 4.28252708959167\\
0.9 4.2429638163508\\
1 4.08342315708289\\
1.1 3.70862204042916\\
1.2 3.03576729890812\\
1.3 2.00358366946856\\
1.4 0.579369357646187\\
1.5 -1.23693676465782\\
1.6 -3.41151373123126\\
1.7 -5.87831858960729\\
1.8 -8.54338876918501\\
1.9 -11.2913816219202\\
2 -13.994195575576\\
2.1 -16.5210069685271\\
2.2 -18.7490473924249\\
2.3 -20.573674166914\\
2.4 -21.9167477411436\\
2.5 -22.7328998591698\\
2.6 -23.0123960977823\\
2.7 -22.780660346634\\
2.8 -22.0953710755295\\
2.9 -21.0410030111655\\
3 -19.7213160762659\\
3.1 -18.2500910873701\\
3.2 -16.7405786415524\\
3.3 -15.2953288860717\\
3.4 -13.9979183841627\\
3.5 -12.9061327322845\\
3.6 -12.0469281836528\\
3.7 -11.4144496793607\\
3.8 -10.9711950870423\\
3.9 -10.6519967283093\\
4 -10.3704138436763\\
4.1 -10.0281352578138\\
4.2 -9.52543190831794\\
4.3 -8.77093142877228\\
4.4 -7.69088860060969\\
4.5 -6.23704924730745\\
4.6 -4.39256322756458\\
4.7 -2.17481259576191\\
4.8 0.364471891209067\\
4.9 3.14189515008181\\
5 6.04844959224601\\
5.1 8.95784305505484\\
5.2 11.7369288717808\\
5.3 14.256779654683\\
5.4 16.4032524261142\\
5.5 18.0861131479366\\
5.6 19.2459895188813\\
5.7 19.8583417881002\\
5.8 19.9343585244696\\
5.9 19.5196703027868\\
6 18.6900805215229\\
6.1 17.5447600820486\\
6.2 16.197140641045\\
6.3 14.7643444409935\\
6.4 13.3570226463222\\
6.5 12.0695922892742\\
6.6 10.9718894555571\\
6.7 10.1033827603784\\
6.8 9.470340081861\\
6.9 9.04661809593262\\
7 8.77657884402986\\
7.1 8.58049250523373\\
7.2 8.36339846883424\\
7.3 8.02498997085566\\
7.4 7.4692669004716\\
7.5 6.6141398846947\\
7.6 5.40019286079838\\
7.7 3.79753373563341\\
7.8 1.81007010253951\\
7.9 -0.523659721730613\\
8 -3.13292982127707\\
8.1 -5.91959856139243\\
8.2 -8.76500407580113\\
8.3 -11.5393931286493\\
8.4 -14.1125212863121\\
8.5 -16.3642721689111\\
8.6 -18.193956082543\\
8.7 -19.5281172003525\\
8.8 -20.3264232352831\\
8.9 -20.5846488894028\\
9 -20.33437227504\\
9.1 -19.6396271355101\\
9.2 -18.5908188213211\\
9.3 -17.2960141181817\\
9.4 -15.8711085543521\\
9.5 -14.4296596212228\\
9.6 -13.0732273237789\\
9.7 -11.882679841566\\
9.8 -10.9108127536143\\
9.9 -10.1780230116043\\
10 -9.6712703710477\\
10.1 -9.34591042234437\\
10.2 -9.13028369675161\\
10.3 -8.93292774492226\\
10.4 -8.65252699067716\\
10.5 -8.18829720599319\\
10.6 -7.44956250151752\\
10.7 -6.36478926927338\\
10.8 -4.88964210033866\\
10.9 -3.0130594140268\\
11 -0.759281894426121\\
11.1 1.8132886192067\\
11.2 4.61519623962334\\
11.3 7.53217246266953\\
11.4 10.4341627822984\\
11.5 13.185517132663\\
11.6 15.655756372581\\
11.7 17.7302023035303\\
11.8 19.3190949625717\\
11.9 20.364570375006\\
12 20.8457518245324\\
12.1 20.7801107072532\\
12.2 20.2205828542804\\
12.3 19.2502275481748\\
12.4 17.974428305028\\
12.5 16.5114799339151\\
12.6 14.9821684223767\\
12.7 13.4993992665035\\
12.8 12.1584094938269\\
12.9 11.0284977410284\\
13 10.1472327924675\\
13.1 9.51759987071184\\
13.2 9.10894123945144\\
13.3 8.8604746724734\\
13.4 8.68715617150909\\
13.5 8.4880007247172\\
13.6 8.15654397375909\\
13.7 7.59185459701477\\
13.8 6.70886343369976\\
13.9 5.44699467413364\\
14 3.7766814104602\\
14.1 1.70389915204422\\
14.2 -0.729499705444842\\
14.3 -3.4486035290629\\
14.4 -6.35024589067252\\
14.5 -9.31009709389521\\
14.6 -12.1919077276517\\
14.7 -14.8583469678168\\
14.8 -17.1822722202014\\
14.9 -19.057288806691\\
15 -20.4062282769274\\
15.1 -21.1873135917024\\
15.2 -21.3971607299846\\
15.3 -21.0701771515892\\
15.4 -20.2753935676625\\
15.5 -19.1100953794111\\
15.6 -17.6906472627367\\
15.7 -16.1421244672246\\
15.8 -14.5863689351733\\
15.9 -13.1309169387892\\
16 -11.8601581670287\\
16.1 -10.8279619401336\\
16.2 -10.0534326769023\\
16.3 -9.52024167081736\\
16.4 -9.17948591582999\\
16.5 -8.95570087556174\\
16.6 -8.75441192794451\\
16.7 -8.47139161571222\\
16.8 -8.00305050772873\\
16.9 -7.25716303995137\\
17 -6.16244163296219\\
17.1 -4.67583385396476\\
17.2 -2.78779807235719\\
17.3 -0.524419764789231\\
17.4 2.05389137085628\\
17.5 4.85624976721983\\
17.6 7.76721290733753\\
17.7 10.6558753933508\\
17.8 13.3867529601159\\
17.9 15.8309066417847\\
18 17.8758310724572\\
18.1 19.4339451968751\\
18.2 20.4493876190327\\
18.3 20.9023138964934\\
18.4 20.8099784974729\\
18.5 20.2244421704722\\
18.6 19.22728060575\\
18.7 17.9224304028346\\
18.8 16.4272913941745\\
18.9 14.8625257481766\\
19 13.3415439627056\\
19.1 11.9603821928149\\
19.2 10.7892018387944\\
19.3 9.86644353741104\\
19.4 9.19658501919927\\
19.5 8.75043228847864\\
19.6 8.46806872007931\\
19.7 8.26501408574554\\
19.8 8.04110530124551\\
19.9 7.69075149577323\\
20 7.1138236801688\\
};
\addplot [
color=red,
solid,
forget plot
]
table[row sep=crcr]{
0 0\\
0.1 -0.467798419057042\\
0.2 -2.39865070339418\\
0.3 -2.84474286783551\\
0.4 -2.39046824432054\\
0.5 -2.94404484168814\\
0.6 -3.4514925999807\\
0.7 -2.92760323003387\\
0.8 -4.07000503890764\\
0.9 -4.46873087281541\\
1 -4.82962362683316\\
1.1 -4.97396189838065\\
1.2 -5.07884697690241\\
1.3 -5.50702663659888\\
1.4 -5.92723559525138\\
1.5 -5.59365942392535\\
1.6 -5.05808079386194\\
1.7 -4.98235054098427\\
1.8 -5.45236819113365\\
1.9 -4.90666992518752\\
2 -4.1824873329696\\
2.1 -3.49723520881087\\
2.2 -4.088944390007\\
2.3 -3.82897188870976\\
2.4 -3.6392910756092\\
2.5 -4.25910285422977\\
2.6 -4.19115416721137\\
2.7 -4.19862213687849\\
2.8 -3.76969704015944\\
2.9 -5.20356758283175\\
3 -4.66358144598747\\
3.1 -5.43921254140034\\
3.2 -5.42888002992362\\
3.3 -5.60204438436234\\
3.4 -5.8138576680059\\
3.5 -6.80819940995074\\
3.6 -6.82199043540703\\
3.7 -7.36335027425326\\
3.8 -6.49377896466413\\
3.9 -7.48664327960544\\
4 -7.30470947784527\\
4.1 -7.70071546314212\\
4.2 -7.9452924886198\\
4.3 -8.01251363507567\\
4.4 -8.59318654844329\\
4.5 -8.07431749604018\\
4.6 -9.09314036717065\\
4.7 -9.25722309019521\\
4.8 -9.33598593147831\\
4.9 -9.74842523275117\\
5 -10.3782966651069\\
5.1 -10.0704516452092\\
5.2 -9.88630908209243\\
5.3 -11.3635800120526\\
5.4 -10.9589170721177\\
5.5 -11.2165077671733\\
5.6 -11.8496655762741\\
5.7 -11.7132917776759\\
5.8 -11.7561489262413\\
5.9 -12.528720687925\\
6 -11.614598931373\\
6.1 -13.4313275377371\\
6.2 -13.4961026631242\\
6.3 -13.0639109303035\\
6.4 -13.6229165732556\\
6.5 -13.6604910365388\\
6.6 -13.9797390912454\\
6.7 -15.0418550063972\\
6.8 -13.8011516252184\\
6.9 -14.5391391566412\\
7 -14.5363277131397\\
7.1 -15.517673262124\\
7.2 -15.8179098319041\\
7.3 -15.5396603060083\\
7.4 -15.7090537571256\\
7.5 -15.7323278996134\\
7.6 -16.4797131623741\\
7.7 -17.0020006139952\\
7.8 -16.5102236945006\\
7.9 -16.5476487982762\\
8 -17.0061039429752\\
8.1 -17.8869451482128\\
8.2 -16.9283601204243\\
8.3 -17.7922217981382\\
8.4 -18.2537886841756\\
8.5 -18.0903188590634\\
8.6 -18.3004129032129\\
8.7 -17.968645802895\\
8.8 -18.2987371194885\\
8.9 -18.521415146232\\
9 -19.0444585398701\\
9.1 -19.0588134370392\\
9.2 -19.3646137343657\\
9.3 -19.1059048057807\\
9.4 -19.9041195627169\\
9.5 -19.6260985199725\\
9.6 -19.2805187935454\\
9.7 -20.3783489006248\\
9.8 -20.4274249786751\\
9.9 -20.0376018186841\\
10 -20.7858731066476\\
10.1 -20.2321626829327\\
10.2 -20.7298503367968\\
10.3 -20.726318650534\\
10.4 -20.8759033550316\\
10.5 -20.9120660846476\\
10.6 -21.4830743354259\\
10.7 -20.2726950242735\\
10.8 -21.3413940274238\\
10.9 -21.8138388610627\\
11 -21.5618742492164\\
11.1 -21.2466438929668\\
11.2 -22.0475945424484\\
11.3 -21.6484190690989\\
11.4 -22.0337147198911\\
11.5 -22.2142282129854\\
11.6 -22.1404007256624\\
11.7 -22.3127480487457\\
11.8 -22.3454968377114\\
11.9 -22.2753204893173\\
12 -23.1053665044307\\
12.1 -23.2642331553501\\
12.2 -22.6966352809475\\
12.3 -22.8865338737025\\
12.4 -22.9318727160533\\
12.5 -22.9552546884606\\
12.6 -23.1132348796598\\
12.7 -23.2998334729861\\
12.8 -23.2833038517233\\
12.9 -23.3801568976187\\
13 -23.0079347089264\\
13.1 -23.3395065517078\\
13.2 -23.6546220147208\\
13.3 -23.5520742822025\\
13.4 -23.3615307234738\\
13.5 -23.6866350109388\\
13.6 -23.5682441177452\\
13.7 -23.796175801724\\
13.8 -24.0248659024167\\
13.9 -24.1743380476462\\
14 -24.2639619731468\\
14.1 -24.1797395557284\\
14.2 -24.020797299527\\
14.3 -24.236324491035\\
14.4 -23.8833763392775\\
14.5 -24.1275343978978\\
14.6 -24.7914221876286\\
14.7 -24.4141348716061\\
14.8 -24.2700417942883\\
14.9 -24.7710672592835\\
15 -25.2201085556125\\
15.1 -24.6267769098265\\
15.2 -24.1870803199076\\
15.3 -24.6679270793369\\
15.4 -24.6909114238828\\
15.5 -25.5190988568946\\
15.6 -24.8315326509913\\
15.7 -24.2478063928042\\
15.8 -24.8434636826976\\
15.9 -25.3398281871499\\
16 -24.5894398419428\\
16.1 -25.0788662047561\\
16.2 -24.8694263610446\\
16.3 -25.28996033173\\
16.4 -25.0684309876354\\
16.5 -25.3168363673881\\
16.6 -25.5398698844739\\
16.7 -24.9342497132735\\
16.8 -25.9569494825808\\
16.9 -25.2152439920697\\
17 -25.5884791928293\\
17.1 -25.8586473235129\\
17.2 -25.5342456258795\\
17.3 -26.0222203468003\\
17.4 -25.2284392122665\\
17.5 -25.6539455753816\\
17.6 -25.3562345371883\\
17.7 -26.0222822023192\\
17.8 -26.0395101254799\\
17.9 -24.7441294038294\\
18 -25.9240183064355\\
18.1 -26.3774442769094\\
18.2 -25.1374355182834\\
18.3 -25.8160546745733\\
18.4 -26.1596626240862\\
18.5 -26.387393293305\\
18.6 -26.0365291905233\\
18.7 -25.899765690425\\
18.8 -26.2333598057959\\
18.9 -26.4821198034276\\
19 -26.0739969401257\\
19.1 -26.5623359642012\\
19.2 -25.8827487059911\\
19.3 -26.0096532000065\\
19.4 -26.484882707462\\
19.5 -25.8422611603658\\
19.6 -26.2301812314003\\
19.7 -25.8419060666701\\
19.8 -26.5460762884569\\
19.9 -26.5281811320565\\
20 -26.2198583377556\\
};
\addplot [
color=red,
solid,
forget plot
]
table[row sep=crcr]{
0 0\\
0.1 -0.729972368134278\\
0.2 -0.938211995070342\\
0.3 -0.150832893264844\\
0.4 0.73816210591532\\
0.5 1.33559054142084\\
0.6 2.13384065054467\\
0.7 3.05394120842153\\
0.8 3.80148811093411\\
0.9 5.19197029384891\\
1 6.90776967313971\\
1.1 8.9367111611957\\
1.2 11.1697921940073\\
1.3 13.5837345837327\\
1.4 16.2636055255677\\
1.5 19.0333453354446\\
1.6 21.3897184104053\\
1.7 23.1577558825876\\
1.8 24.5043166390946\\
1.9 25.5129936387502\\
2 25.6101660254395\\
2.1 24.6919281654743\\
2.2 22.913683007573\\
2.3 20.8926148337882\\
2.4 18.321989482301\\
2.5 15.4367507746101\\
2.6 12.7098219241763\\
2.7 9.94528400998073\\
2.8 7.27928852807223\\
2.9 4.77652484986314\\
3 3.25216886424356\\
3.1 1.87824757496697\\
3.2 1.21850944047431\\
3.3 0.854638237712874\\
3.4 0.789681656217386\\
3.5 0.999745827923668\\
3.6 1.65162474188791\\
3.7 2.17623741623463\\
3.8 2.58164006700476\\
3.9 2.23204082732998\\
4 1.86536436801042\\
4.1 0.92844046466295\\
4.2 -0.294297249361071\\
4.3 -1.8459289939856\\
4.4 -3.6719448102851\\
4.5 -5.49383096071881\\
4.6 -7.55652771342969\\
4.7 -9.01666241025075\\
4.8 -10.1393932074183\\
4.9 -10.8095817493686\\
5 -10.7442622581869\\
5.1 -9.87093264478734\\
5.2 -8.63731681840957\\
5.3 -6.87559356780926\\
5.4 -4.02829771376326\\
5.5 -1.09327221023653\\
5.6 2.12492326539687\\
5.7 5.5613892904092\\
5.8 8.6883193797844\\
5.9 11.5038155229093\\
6 14.0811492569696\\
6.1 15.7346278559532\\
6.2 17.597696094178\\
6.3 18.763759958507\\
6.4 19.11839059567\\
6.5 19.182996040423\\
6.6 18.8454029306443\\
6.7 18.456782160124\\
6.8 18.3332280970071\\
6.9 17.6671181116927\\
7 17.4860340876701\\
7.1 17.6044321673099\\
7.2 18.4919330287538\\
7.3 19.7511094646062\\
7.4 21.0878964038426\\
7.5 22.6229331754555\\
7.6 24.2304998906827\\
7.7 26.0996697649869\\
7.8 27.8636596708019\\
7.9 28.9014274588182\\
8 29.372956893079\\
8.1 29.4260898294915\\
8.2 29.0446520641517\\
8.3 27.4668859853507\\
8.4 25.5865220225405\\
8.5 23.2359580499218\\
8.6 20.2647440190525\\
8.7 16.9504262636402\\
8.8 13.2577478300577\\
8.9 9.67971715611899\\
9 6.36173959503883\\
9.1 3.54603004279074\\
9.2 1.12882696007356\\
9.3 -0.729889195240013\\
9.4 -2.18923768225477\\
9.5 -2.8597270297667\\
9.6 -3.35348583005388\\
9.7 -3.67971572286091\\
9.8 -3.34973791476899\\
9.9 -3.05878033913767\\
10 -3.05848953407531\\
10.1 -3.02791403566863\\
10.2 -3.60494799471347\\
10.3 -4.36785755628212\\
10.4 -5.56424559355408\\
10.5 -7.09565658233803\\
10.6 -8.8936574976137\\
10.7 -10.7469882886459\\
10.8 -13.1906523312369\\
10.9 -15.0043839787243\\
11 -16.3828244772991\\
11.1 -17.5376581240963\\
11.2 -18.2778311023524\\
11.3 -18.0762842967618\\
11.4 -17.4037212558596\\
11.5 -15.8936730123365\\
11.6 -13.7307733874425\\
11.7 -11.1125808666473\\
11.8 -8.07018451148397\\
11.9 -4.84108883532046\\
12 -1.56761734626326\\
12.1 1.95606773503273\\
12.2 5.16580525498546\\
12.3 7.64909396232089\\
12.4 9.67202023536231\\
12.5 11.1164709855925\\
12.6 11.994540719032\\
12.7 12.4272469595842\\
12.8 12.505179669299\\
12.9 12.2625933065055\\
13 11.8571166311192\\
13.1 11.2820022542973\\
13.2 11.0263036431451\\
13.3 11.1602500951676\\
13.4 11.5521742224412\\
13.5 12.2411704809599\\
13.6 13.4331404786446\\
13.7 14.8970358232218\\
13.8 16.707685728559\\
13.9 18.7201904022771\\
14 20.7300084429684\\
14.1 22.5259272736406\\
14.2 23.852170691472\\
14.3 24.5620253172345\\
14.4 24.6756754956408\\
14.5 23.8917315815655\\
14.6 22.5192567088941\\
14.7 20.708828661109\\
14.8 18.0583427079447\\
14.9 14.8783452939483\\
15 11.635762670072\\
15.1 8.38700367108435\\
15.2 4.82206186695184\\
15.3 1.27390531432444\\
15.4 -1.68508513542748\\
15.5 -4.08338438427879\\
15.6 -5.59465747346192\\
15.7 -6.93463887521084\\
15.8 -7.97129769337635\\
15.9 -8.2065139714041\\
16 -7.92480548191584\\
16.1 -7.77460987955929\\
16.2 -7.36807796714253\\
16.3 -7.1497878438665\\
16.4 -7.00781964760461\\
16.5 -7.32099635631766\\
16.6 -7.92466988833476\\
16.7 -8.92779750205301\\
16.8 -10.5623739224217\\
16.9 -12.0941365677707\\
17 -14.1501652567448\\
17.1 -16.0378833943\\
17.2 -17.7009907103436\\
17.3 -19.189737196876\\
17.4 -20.0733081350514\\
17.5 -20.7452916869947\\
17.6 -20.580037745414\\
17.7 -19.8002052465255\\
17.8 -17.9928925308596\\
17.9 -15.6611027323229\\
18 -13.3401820759991\\
18.1 -10.0221132418112\\
18.2 -6.39099306725494\\
18.3 -3.2971684318309\\
18.4 -0.00293718994603434\\
18.5 3.17016731068152\\
18.6 5.9762902044073\\
18.7 8.06523044887685\\
18.8 9.53832070162705\\
18.9 10.620705664051\\
19 11.2568246922991\\
19.1 11.3047560243567\\
19.2 11.2286315217499\\
19.3 10.6742018878794\\
19.4 10.2226072178691\\
19.5 10.0809276570797\\
19.6 9.90575744059559\\
19.7 10.2134651080062\\
19.8 10.7773347386006\\
19.9 12.0718447397411\\
20 13.653648352172\\
};
\addplot [
color=red,
solid,
forget plot
]
table[row sep=crcr]{
0 0\\
0.1 1.79358319433153\\
0.2 9.48741356465934\\
0.3 11.1708667478066\\
0.4 9.16495784068533\\
0.5 11.2053635064728\\
0.6 13.0864419589864\\
0.7 10.8297805991338\\
0.8 15.3036378857822\\
0.9 16.7593756881858\\
1 17.9528626225631\\
1.1 18.0912832153493\\
1.2 17.8164556256657\\
1.3 18.5291659153331\\
1.4 18.8355804348307\\
1.5 15.6890666046167\\
1.6 11.3181957823795\\
1.7 8.45132178852938\\
1.8 7.51361099530632\\
1.9 2.2854564392061\\
2 -3.75657090161228\\
2.1 -9.59405547632656\\
2.2 -10.0814311976351\\
2.3 -13.6855097970385\\
2.4 -16.6248885122933\\
2.5 -15.856512044337\\
2.6 -17.3987530195078\\
2.7 -18.1955041223808\\
2.8 -20.3398142532527\\
2.9 -14.6121711975124\\
3 -16.5891334312904\\
3.1 -13.1352221060111\\
3.2 -12.805149353627\\
3.3 -11.8037886481349\\
3.4 -10.7827878322113\\
3.5 -6.79186644012375\\
3.6 -6.98157665999896\\
3.7 -5.28802968325028\\
3.8 -9.50046506643558\\
3.9 -6.34450887950019\\
4 -7.95766481884041\\
4.1 -7.19833313742764\\
4.2 -6.90372936197379\\
4.3 -7.0961222291029\\
4.4 -4.90517292529854\\
4.5 -6.78064339073025\\
4.6 -2.05765586468886\\
4.7 -0.387441766275737\\
4.8 1.25523161939463\\
4.9 4.48734643932271\\
5 8.73444555381666\\
5.1 9.19924842021374\\
5.2 10.033640883561\\
5.3 17.3422824145139\\
5.4 16.7121731381326\\
5.5 18.2778601837903\\
5.6 20.8492983217702\\
5.7 19.7865054291959\\
5.8 18.9198391147195\\
5.9 20.5278020868774\\
6 14.9319037769107\\
6.1 20.0516136198936\\
6.2 17.9390135008287\\
6.3 13.7141356985134\\
6.4 13.5057462978013\\
6.5 11.3289527032206\\
6.6 10.4921291935464\\
6.7 12.9080657897977\\
6.8 6.27825201774566\\
6.9 7.8387455143076\\
7 6.60817862784738\\
7.1 9.44749892191022\\
7.2 9.53635951452177\\
7.3 7.15780896765469\\
7.4 6.3686242597404\\
7.5 4.71442097083771\\
7.6 5.64953492745654\\
7.7 5.30755085883063\\
7.8 0.47834154682488\\
7.9 -2.5678679365051\\
8 -4.15253536730122\\
8.1 -4.16777385375159\\
8.2 -11.6497537474486\\
8.3 -11.7007194003994\\
8.4 -13.1282650968514\\
8.5 -16.7477807327753\\
8.6 -18.4425825828234\\
8.7 -21.8223448357112\\
8.8 -21.9838301919525\\
8.9 -22.0113542240497\\
9 -20.3003146243932\\
9.1 -20.1823435310191\\
9.2 -18.5309132314929\\
9.3 -18.9093544081507\\
9.4 -14.8926281782973\\
9.5 -15.1963788438938\\
9.6 -15.8621974587324\\
9.7 -10.8440610204806\\
9.8 -10.2423748655469\\
9.9 -11.6487288957864\\
10 -8.67492192093109\\
10.1 -11.1086595553166\\
10.2 -9.3933725016367\\
10.3 -9.68991599262625\\
10.4 -9.27337459092307\\
10.5 -9.11841831953975\\
10.6 -6.52860493013888\\
10.7 -10.7844393544505\\
10.8 -5.46081950688318\\
10.9 -2.10854383698087\\
11 -1.30901726189146\\
11.1 -0.463474279463271\\
11.2 5.11821456979884\\
11.3 5.98925563560907\\
11.4 10.0124231681476\\
11.5 13.0761724479046\\
11.6 14.8432881702864\\
11.7 17.2204124250217\\
11.8 18.5722217139163\\
11.9 18.9903531633209\\
12 22.500807872337\\
12.1 22.7723545658011\\
12.2 19.6101207424105\\
12.3 19.0916149718714\\
12.4 17.6985216986888\\
12.5 16.0389896210708\\
12.6 14.8674896859504\\
12.7 13.866529071298\\
12.8 12.1853452654685\\
12.9 11.1672910790216\\
13 8.50780119811977\\
13.1 8.94512770939561\\
13.2 9.54787717812203\\
13.3 8.62062389655523\\
13.4 7.40083601647799\\
13.5 8.23703650616293\\
13.6 7.1692371471988\\
13.7 7.27734810232218\\
13.8 7.0787267499395\\
13.9 6.18255891756277\\
14 4.63666276777716\\
14.1 1.98385733654738\\
14.2 -1.33337539428023\\
14.3 -3.41793278292308\\
14.4 -7.9692415437784\\
14.5 -10.1617195921308\\
14.6 -10.5518200093928\\
14.7 -14.9095642304168\\
14.8 -17.9839209189595\\
14.9 -17.9952139287169\\
15 -17.6892349576419\\
15.1 -21.0404083740843\\
15.2 -23.2248033044934\\
15.3 -21.1476632044584\\
15.4 -20.4259831679912\\
15.5 -16.0863893346864\\
15.6 -17.6206751801751\\
15.7 -18.6300177368108\\
15.8 -14.859277824204\\
15.9 -11.5702452981809\\
16 -13.5057025805912\\
16.1 -10.6873722179571\\
16.2 -10.9376172862184\\
16.3 -8.88954478815913\\
16.4 -9.61935823486635\\
16.5 -8.57428070643753\\
16.6 -7.65206814014683\\
16.7 -9.99835846018119\\
16.8 -5.57854658542788\\
16.9 -7.96691560500325\\
17 -5.51076014866262\\
17.1 -3.06889223861746\\
17.2 -2.63013063923842\\
17.3 1.44737966721895\\
17.4 0.661670002013775\\
17.5 5.01317908771081\\
17.6 6.57311688784186\\
17.7 12.0088428270778\\
17.8 14.6738111892318\\
17.9 11.7330566259815\\
18 18.3822896454535\\
18.1 21.6463551487687\\
18.2 17.5163256518527\\
18.3 20.5598686351908\\
18.4 21.7525446668724\\
18.5 21.9933496104498\\
18.6 19.4816440206767\\
18.7 17.5184858041074\\
18.8 17.2639889652423\\
18.9 16.6011984315754\\
19 13.3242672395173\\
19.1 13.7952716327019\\
19.2 9.76088319990677\\
19.3 9.21658758686943\\
19.4 10.3362642220145\\
19.5 7.16163405710891\\
19.6 8.29603104133853\\
19.7 6.3995103898492\\
19.8 8.9055350680091\\
19.9 8.38445641333482\\
20 6.45479991936009\\
};
\end{axis}
\end{tikzpicture}%